\documentclass[12pt,preprint]{aastex}

\slugcomment{\it Accepted by ApJ}
\shorttitle{Mid-Infrared Counterparts of Submm Galaxies}
\shortauthors{Ashby et al.}

\def\deg{\ifmmode {^{\circ}}\else {$^\circ$}\fi}

\def\ergcm2s{\ifmmode {\rm\,erg\,cm^{-2}\,s^{-1}}\else
     ${\rm\,erg\,cm^{-2}\,s^{-1}}$\fi}
\def\spose#1{\hbox to 0pt{#1\hss}}
\def\simlt{\mathrel{\spose{\lower 3pt\hbox{$\mathchar"218$}}
      \raise 2.0pt\hbox{$\mathchar"13C$}}}
\def\simgt{\mathrel{\spose{\lower 3pt\hbox{$\mathchar"218$}}
      \raise 2.0pt\hbox{$\mathchar"13E$}}}

\begin{document}

\newcommand{\NGC}{\mbox{\protect\small NGC\hspace{.15em}\normalsize}}
\newcommand{\etal}{{et al.}~}
\newcommand{\Msun}{$M_\odot$}
\newcommand{\Lsun}{$L_\odot$}
\newcommand{\irac}{{\sl IRAC}}
\newcommand{\sst}{{\sl Spitzer Space Telescope}}

\title{Mid-Infrared Identifications of SCUBA Galaxies \\
in the CUDSS 14-Hour Field \\ with the \sst}

\author{
M. L. N. Ashby\altaffilmark{1},
S. Dye\altaffilmark{2},
J.-S. Huang\altaffilmark{1},
S. Eales\altaffilmark{2},
S.~P. Willner\altaffilmark{1},
T. M. A. Webb\altaffilmark{3},
P. Barmby\altaffilmark{1},
D. Rigopoulou\altaffilmark{4},
E. Egami\altaffilmark{5},
H. McCracken\altaffilmark{6},
S. Lilly\altaffilmark{7},
S. Miyazaki\altaffilmark{8},
M. Brodwin\altaffilmark{9},
M. Blaylock\altaffilmark{5},
J. Cadien\altaffilmark{5},
and 
G. G. Fazio\altaffilmark{1} 
}

\altaffiltext{1}{Harvard-Smithsonian Center for Astrophysics, 60 Garden Street, 
Cambridge, MA  02138;
mashby, jhuang, swillner, pbarmby, gfazio@cfa.harvard.edu} 

\altaffiltext{2}{School of Physics and Astronomy, Cardiff University, 5 The Parade, Cardiff, CF24 3YB, UK}

\altaffiltext{3}{McGill University Department of Physics,
Rutherford Physics Building, 3600 rue University,
Montr\'eal, Qu\'ebec,
Canada H3A 2T8}

\altaffiltext{4}{Department of Astrophysics, DWB, Oxford University, Keble Rd, 
Oxford, OX1 3RH, UK}

\altaffiltext{5}{Steward Observatory, University of Arizona, 933 North Cherry Avenue,
Tucson, AZ 85721}

\altaffiltext{6}{Institute d'Astrophysique, 98bis, Bd Arago - 75014 Paris, France}

\altaffiltext{7}{
Institute of Astronomy, Swiss Federal Institute of Technology Zurich, ETH Hoenggerberg Campus
Physics Department, HPF G4.1, CH-8093 Zurich, Switzerland}

\altaffiltext{8}{Subaru Telescope, National Astronomical Observatory of Japan, 
650 North A'ohoku Place, Hilo, HI 96720}

\altaffiltext{9}{Jet Propulsion Laboratory, California Institute of
Technology, Mail Stop 169-506, Pasadena, CA 91109}

\begin{abstract}
We identify 17 possible 8.0~$\mu$m-selected counterparts to the submillimeter 
galaxies in the CUDSS 14$^{\rm h}$ field, derived from deep imaging carried
out with the IRAC and MIPS instruments aboard the {\sl Spitzer Space Telescope}.  
Ten of the 17 counterparts are not the same as those previously identified 
at shorter wavelengths.  We argue that 8.0~$\mu$m selection offers a better 
means for identifying counterparts to submillimeter galaxies than near-infrared or
optical selection.  Based on the panchromatic SEDs, most counterparts appear 
to be powered by ongoing star formation.  Power-law fits to the SEDs suggest
that five objects in the 8.0~$\mu$m-selected counterpart sample harbor 
dominant AGNs; a sixth object is identified as a possible AGN.
The 3.6-8.0~$\mu$m colors of the infrared-selected counterparts are 
significantly redder than the general IRAC galaxy population in the
CUDSS 14$^{\rm h}$ field.

\end{abstract}


\section{Introduction}

Observations at 850\micron, initially with the Submillimeter Common
User Bolometer Array (SCUBA) and subsequently with the Max Planck
Millimeter Bolometer (MAMBO), have revealed galaxies that are strong
sources of dust emission but faint at visible wavelengths (Smail,
Ivison \& Blain 1997, Hughes \etal\ 1998, Barger \etal\ 1998, Eales
etal\ 1999, Clements \etal\ 2004). 

The SCUBA galaxy populations are important because they
seem to represent the most luminous star-forming galaxies during the
epoch when star formation rates were highest.  The space density of 
submillimeter galaxies is orders of magnitude greater than that of 
$10^{13}$\Lsun\ galaxies at low redshift (Blain \etal 2002), 
implying that strong evolution has greatly altered
the submillimeter galaxy populations since $z=2$.  Moreover, some or
all of the SCUBA population may be the progenitors of today's massive
elliptical galaxies.  The stars in nearby elliptical galaxies and in
the bulges of disk galaxies are old and tend to be metal rich. 
At high redshifts, there ought to be a significant population of
progenitor galaxies with star formation rates (SFRs) high enough to
produce today's population of ellipticals and bulges (Lilly \etal\
1999).  No such population has been detected in deep optical surveys,
but the SCUBA galaxies, or some of them, could represent the
long-sought progenitors.
Understanding these so-called ``SCUBA galaxies'' and their place 
in cosmic evolution, however, requires complementary observations at 
other wavelengths and in particular some way to determine redshifts.

Observing SCUBA galaxies in visible light is difficult.  The
dust responsible for the submillimeter emission obscures the primary
luminosity sources, making the SCUBA galaxies quite faint at visible
wavelengths.  Even identifying the correct visible counterpart to a given
SCUBA galaxy is difficult because of the large beamsize (14\arcsec\
FWHM at 850~$\mu$m) of the James Clerk Maxwell Telescope and other
submillimeter facilities. The unavoidable result is that at visible
and even near-infrared wavelengths, there are numerous potential
counterparts in the SCUBA beam, and there is no easy way to decide
which of the candidates is the true counterpart.

One successful approach to finding counterparts has been to use radio
interferometric observations to refine the SCUBA positions and thus
select a specific visible object.  Chapman \etal\ (2003, 2005), for
example, have used this approach and have shown that radio-identified 
SCUBA galaxies have a median spectroscopic redshift of $\langle z
\rangle=2.4$.  However, about 1/3 of SCUBA sources are not detectable
with current radio observations (Chapman \etal\ 2005) and could
represent a distinct population.  It is therefore worthwhile seeking
alternate ways to identify counterparts.


This contribution describes a new program to recover submm-selected 
galaxies in the mid-infrared, so as to exploit the advantages 
this passband offers for better identifications of the counterparts
to these important objects.
Our submm sample is comprised of 23 SCUBA-detected objects in the
14$^{\rm h}$ field of the Canada-UK Deep Submillimetre Survey (CUDSS;
Lilly \etal 1999, Eales \etal 2000, Webb \etal 2003).
We base our infrared counterpart search on the very deep 8.0~$\mu$m 
{\sl Spitzer} mosaic of the 14$^{\rm h}$ field.
Although the {\sl Spitzer} observations (described below) include 
coextensive observations at a number of passbands from 3.6 to 160~$\mu$m,
we use sources drawn specifically from the 8.0~$\mu$m 
mosaic, for the following reasons.  

First, 8.0~$\mu$m observations sample rest-frame 
{\sl K}-band emission at typical SCUBA galaxy redshifts $\langle z \rangle=2.4$.
They are therefore much less affected than shorter-wavelength observations 
by extinction due to the dust known to be abundant in galaxies that
are actively forming stars (Schmitt \etal 2006).  
This rest-frame {\sl K}-band emission is also a good 
proxy for stellar mass (Cole \etal 2001); it is therefore more sensitive to
total star formation history than shorter-wavelength observations 
that are relatively more affected by recent star formation.

Second, high-redshift galaxies like those selected at submm wavelengths 
typically have red infrared colors that are well-suited to detection 
at 8.0~$\mu$m (Egami \etal 2004, Huang \etal 2004, Frayer \etal 2004b), 
or at least they have red optical and near-infrared colors 
(Pope \etal 2005, Frayer \etal 2004a, Borys \etal 2004, Ivison \etal 2002).
Lockman Hole galaxies were detected in 450 and 850~$\mu$m 
SCUBA maps with stacking analyses at 5.8 and 8.0~$\mu$m
by Serjeant \etal (2004) but not at 3.6 or 4.5~$\mu$m.
In addition, although the 3.6 and 4.5~$\mu$m 
{\sl Spitzer} observations in the 14$^{\rm h}$ field are somewhat more 
sensitive than the 8.0~$\mu$m mosaic, they tend to detect more of the
blue, relatively numerous, low-redshift galaxies that are unrelated 
to the SCUBA-selected population.  
The resulting contamination problem is of course more severe
at shorter wavelengths, e.g., {\sl K}-band.  
At 8.0~$\mu$m we avoid to a great degree
the problems of false associations that affect these other bands, without
missing sources: all potential 8.0~$\mu$m counterparts we considered 
are also detected in the 3.6~$\mu$m mosaic.

Third, the 8.0~$\mu$m mosaics themselves are very deep -- 
down to a 5$\sigma$ point source sensitivity of 5.8~$\mu$Jy.
We used two very simple models of star formation history to
calculate that this depth is easily sufficient to detect typical
SCUBA galaxies.  Both models (one a constant star formation rate 
model, the other a burst) give approximately equivalent results and 
suggest that
at redshifts of $z=2-3$ the 8.0~$\mu$m mosaics will detect star
formation rates of $120\pm30$ and $190\pm40$\Msun\ yr$^{-1}$, 
respectively. 
This is comfortably below the rates found for submm galaxies 
(up to roughly $10^3$\Msun\ yr$^{-1}$, e.g., Smail \etal 2004, 
Yun \& Carilli 2002, Lilly \etal 1999), so we can expect {\it a priori} 
that their 8.0~$\mu$m counterparts are detected in the {\sl Spitzer} data.

This paper describes the {\sl Spitzer} observations, the 8.0~$\mu$m counterpart
identification procedure, and the inferences to be drawn from
the panchromatic SEDs of infrared-selected SCUBA counterparts.
A companion paper (Dye \etal 2006) describes the inverse process,
whereby we determine the average submillimeter emission at the
positions of the {\sl Spitzer} sources found in the field.

The next Section describes the {\sl Spitzer}/IRAC and MIPS observations and
presents infrared images at the known locations of the SCUBA galaxies.
We then outline the scheme used to identify the most-likely infrared counterparts
of the submillimeter galaxies in Section~\ref{sec:ids}.  The
infrared properties of the likely counterparts are analyzed in
Section~\ref{sec:discussion}.

\section{Observations and Data Reduction}

\subsection{The IRAC Observations}

The IRAC (Fazio \etal 2004) observations were carried out as part of 
Spitzer Guaranteed Time Observing program number 8 to deeply image the 
Extended Groth Strip (EGS).  The first visit to the field was in 2003 December.  
The field was observed again during a second epoch in 2004 June at a 
different position angle.
The IRAC exposures (each of which covered a 5\farcm12$\times5$\farcm12 field 
of view with $256\times256$ pixels $1\farcs2$ in size) 
totaled 52 dithered 200~s exposures at 3.6, 4.5, and 5.8~$\mu$m, together with
208 dithered 50~s exposures taken concurrently at 8.0~$\mu$m at all positions
in a $2^\circ \times 10^\prime$ map.  Each dither cluster was separated from its neighbor by
290\arcsec, slightly less than the IRAC field of view.
Because the field was observed at position angles differing by
roughly $180^\circ$ during the
two epochs it was straightforward to remove most of the well-known instrumental
artifacts (multiplexer bleed, banding) during mosaicing.
The resulting point-source sensitivity 
was 24.0, 24.0, 21.9, and 22.0 mag (AB, 5$\sigma$), 
in the 3.6, 4.5, 5.8, and 8.0~$\mu$m bands, respectively.  
This corresponds to flux density levels
of 0.9, 0.9, 6.3, and 5.8~$\mu$Jy in the four IRAC bands.
All magnitudes given in this work are on the AB system.  

Sources in the IRAC mosaics were identified and extracted using SExtractor
(Bertin \& Arnouts 1996).  The software was configured with a 2.5$\sigma$
detection threshold and a minimum object area of at least 5 pixels.
The result was a band-merged photometric catalog of objects 
for all four IRAC wavelengths; photometry was performed using 3\arcsec\ 
diameter apertures centered at the positions computed at 3.6~\micron.
All sources here are pointlike in the IRAC bands.  To permit comparison
with total magnitudes in the other bands, appropriate 
corrections were applied to obtain total magnitudes from the aperture 
magnitudes.  The density of 8.0~$\mu$m sources brighter than 5.8~$\mu$Jy 
is roughly 11 per square arcminute.

\subsection{The MIPS Observations}

The MIPS (Rieke et al. 2004) observations of the EGS were carried out on 2004
June 19 and 20.  The MIPS 24~$\mu$m channel ($\lambda=23.7 \mu$m;
$\Delta\lambda = 4.7 \mu$m) used a 128$\times$128 Si:As
array with a pixel scale of 2\farcs55 pixel$^{-1}$, providing a
field of view of $5\farcm4\times$5\farcm4.  
The 70~$\mu$m channel ($\lambda=71.4 \mu$m;
$\Delta\lambda = 19 \mu$m) uses a 32$\times$32 Ge:Ga array with a
pixel scale of 9\farcs98 pixel$^{-1}$, but one half of the array is
not usable due to a high noise caused by a cabling problem.
The effective 70~$\mu$m array size is therefore 32 $\times$16 pixels,
providing a field of view of 5\farcm2$\times$2\farcm6.

The area was observed in scan map mode at slow rate with scan 
legs 2\fdg4 long.  This results in 
an integration time of 100 sec pixel$^{-1}$ per scan pass 
(10 frames$\times$10 s) at both 24 and 70~$\mu$m.  In scan mode 
images at these two wavelengths are obtained simultaneously.
Pairs of scan map observations, with a cross-scan offset of 
one full array (296\arcsec), were executed
6 times, with an offset of 21\arcsec\ between successive pairs.  
The effective integration time is therefore $\sim$1500~s (12 passes) at
24~$\mu$m for locations near the long centerline of the strip,
decreasing to $\sim$700~s 5~\arcmin\ from the centerline.
At 70~$\mu$m, the integration time is only half as much 
due to the loss of half of the detector array.
The final scan maps cover an area 2.4\degr$\times$10\arcmin\ with an
integration time $>700$ sec pixel$^{-1}$. A smaller 2\degr$\times$6\arcmin\ strip 
in the center, encompassing most of the CUDSS field, was covered with an 
integration time of 1300--1600 seconds.

The data were reduced and mosaiced with the MIPS Data Analysis Tool
(Gordon \etal 2004).  The 24~$\mu$m scan images
were resampled and mosaiced at half the original instrument
pixel scale (1\farcs25) while at 70~$\mu$m the original pixel
scale was preserved.
The point spread functions in the mosaics have FWHMs 
of 6\arcsec\ and 18\arcsec\ at 24 and 70~$\mu$m, respectively.  
The resulting 5$\sigma$ point source sensitivities are respectively
70~$\mu$Jy and 7 mJy at 24 and 70~$\mu$m.

\subsection{Observations at Other Wavelengths}

We have drawn from several pre-existing photometric catalogs in order 
to construct SEDs spanning as wide a wavelength range as possible
for our counterparts.  For the U, B, V, and I band photometry, we used the 
Canada-France Deep Field Survey (McCracken \etal 2001, hereafter CFDF)
survey data.  CFDF magnitudes are calculated within 2.5\arcsec\ diameter
apertures.  The 3$\sigma$ CFDF limiting magnitudes are 27.71, 26.23, 25.98, and 25.16 
mag in U, B, V, and I, respectively.
At R we use the very deep image acquired at the Subaru telescope
with {\sl Suprime}; our R-band photometry has a $5\sigma$ limit of
26.6 mag (Miyazaki 2006, private communication).
For z-band we use the catalog constructed by Brodwin \etal (2006), 
converting Vega to AB magnitudes by adding a constant 0.54 mag.
The limiting magnitude of the z-band data is 25.0 ($3\sigma$), within
2.5\arcsec\ diameter apertures.
The $K$-band photometry data are drawn from the observations described 
by Webb \etal (2003), converting to AB magnitudes using $K_{AB} = K_{VEGA} + 1.91$.
$K$-band magnitudes were measured within 3\arcsec\ apertures.
Because the $K$-band data were acquired at two facilities, the depth 
is nonuniform but it generally extends down to $\sim23$ mag ($3 \sigma$).
All visible and near-infrared sources considered here are small compared 
to the apertures used; we therefore treat the aperture magnitudes 
as total magnitudes.

Two of the submillimeter galaxies (14.13 and 14.18) have {\sl Infrared Space
Observatory} (ISO) 15~$\mu$m fluxes reported by Flores \etal (1999).  
One of these (14.13) was observed using {\sl Spitzer's}
{\sl Infrared Spectrograph} (IRS) in low-resolution mode by Higdon \etal (2004), 
who also report a 16~$\mu$m flux density from the IRS Peakup Imager.
Finally, we have also made use of the Chandra X-ray Observatory (CXO) 
source catalog compiled by Nandra \etal (2005). 
A montage showing our observations at R, 3.6, 4.5, 5.8, 8.0, and 24~$\mu$m
at the locations of each of the 23 SCUBA sources in 
the 14$^{\rm h}$ field is provided in Figure~\ref{fig:montage}.

\section{The Mid-Infrared Identifications}
\label{sec:ids}

\subsection{The Identification Procedure}

To avoid spurious sources we restrict our investigation to portions 
of the 14$^{\rm h}$ field covered by at least ten 8~$\mu$m IRAC exposures.
With this constraint the IRAC coverage allows counterpart
identifications for 20 of the 23 SCUBA galaxies in this field.  

We identify IRAC counterparts of the SCUBA sources using a two-part
approach, consisting first of a proximity search of the
IRAC 8.0~$\mu$m source catalog, followed by an
examination of the infrared SEDs of all possible matches.
We first use a metric based on 8.0~$\mu$m brightness and 
positional coincidence to select likely counterparts.  
We restrict the search to IRAC objects within 10\arcsec\ 
of each SCUBA source, for several reasons.  The primary 
motivation is to adopt a (conservative) value slightly larger
than the $\sim8$\arcsec\ uncertainty in the SCUBA positions,
so as not to miss any true counterparts that might lie at
relatively large separations, while simultaneously avoiding 
such a large value that we include a large number of 
unassociated foreground objects in the search.  
We find that a 10\arcsec\ search radius often selected just one 
(or at most three) candidates, and was therefore well-matched to our
IRAC source density.  Based on the source surface density 
down to the 5$\sigma$ limit of our survey, on average one 
IRAC 8~$\mu$m source should randomly reside within each 
10\arcsec\ radius search area.

Borys \etal (2004) found 11 counterparts within 7\arcsec\ of 
19 submillimeter objects in the Hubble Deep Field North, implying
a substantial fraction may lie beyond this radius.
Similarly, Wang \etal (2004) identified five of their 20 X-ray counterparts
and two of their six mid-infrared counterparts to submillimeter galaxies in 
the same field at separations greater than 7\arcsec.
By contrast Lilly \etal\ (1999) found at 
least 90\% of the visible-wavelength counterparts to 
the submillimeter galaxies in the CUDSS 14$^{\rm h}$ field
are present within 7\arcsec, but their analysis includes 
only six sources.  Our 10\arcsec\ criterion is also 
safely larger than the mostly-likely average 850~$\mu$m-infrared 
offset of 2\arcsec\ determined from the stacking analysis discussed 
by Dye \etal (2006).

All potential counterparts are listed in Table~\ref{8micids}.
In seven cases (see below) a single, obvious best counterpart was 
identified.  In instances where the identifications were not clear-cut, 
we applied a second criterion based on infrared color to refine 
the identifications.  Specifically, we made use of $K$-band photometry 
from Webb \etal (2003) as well as our own IRAC
photometry to discriminate the most likely counterpart among
multiple candidates.  In doing so, we made one important assumption.  
We {\sl a priori} assumed, based on spectroscopic studies (Chapman \etal 2003, 
2005), that the true counterparts must reside at high redshift and 
that the SEDs must be consistent with this fact.  
We applied color 
criteria developed by Huang \etal (2004; see their Table~2), who 
use the $K$-to-IRAC colors to constrain a galaxy's redshift into local, 
intermediate, and high-redshift regimes.  Those criteria are summarized as
follows:  a blue $K - [3.6]$ color corresponds to redshifts $z<0.6$. 
Red $K - [3.6]$ color is consistent with either an intermediate ($0.6 < z < 1.3$)
or high redshift ($z > $1.3), depending on whether the mid-infrared $[3.6] - [4.5]$ 
color is blue or red, respectively.  In Section~\ref{ssec:notes} we make frequent
use of this technique to select the best (highest-redshift) candidate among multiple 
possibilities.  Details of the identifications for 
these 20 accessible sources are provided in the following Section.  

\subsection{Individual IRAC Identifications}
\label{ssec:notes}

Seven sources have secure identifications: 
14.1, 14.3, 14.13, 14.17, 14.18, 14.19, and 14.23.
Of these, all but 14.17 and 14.23 have relatively small position uncertainties.
We have possible identifications for ten others: 14.2, 14.4, 14.7, 14.8, 14.10, 14.12,
14.14, 14.15, 14.20, and 14.22.  Because their infrared colors and R-band 
morphologies are not those of dusty, high-redshift galaxies,
we regard both potential 8.0~$\mu$m identifications for 14.11 as dubious 
and do not include it in subsequent analysis.  Likewise 
14.6 and 14.21 which have no significant 8.0~$\mu$m counterpart 
within 10\arcsec\ of the SCUBA positions.

\subsubsection{Sources with Radio Identifications}

Objects 14.1, 14.3, 14.13, and 14.18 have 1.4 GHz positions available
from Fomalont \etal (1991).  A recalibration of the VLA data by M.~Yun
improved the effective beamwidth of this dataset to 4\arcsec\ FWHM 
and resulted in a radio identification for 14.19 (Webb \etal 2003).

{\it CUDSS 14.1}. Our single 8.0~$\mu$m identification 
is coincident with both the Webb \etal (2003) $K$-selected candidate and the
I-selected candidate from Lilly \etal (1999) within the positional uncertainties.  
There is also a MIPS 24~$\mu$m source at this position.

{\it CUDSS 14.3}. Our only 8.0~$\mu$m candidate is coincident with the 
Webb \etal (2003) candidate as well as the position observed by 
Chapman \etal 2005, who used Keck to obtain a spectroscopic redshift of $z=1.139$
for this object.  This object is also an obvious MIPS 24~$\mu$m source.

{\it CUDSS 14.13}.  For this object the best candidates at $K$ and 8.0~$\mu$m agree within 
the positional uncertainties.  This object is also 
a strong MIPS 24~$\mu$m source and was detected 
by ISO at 7 and 15~$\mu$m at the same position.  This is not surprising, because it is by 
far the brightest of all our IRAC candidates.  This object was also observed with the
Infrared Spectrograph on the {\sl Spitzer Space Telescope} (Higdon \etal 2004) 
and exhibits a continuum slope consistent with a Seyfert 2 nucleus and no PAH emission.
Our best position is coincident with that observed by Chapman \etal (2005), 
who used Keck to obtain a spectroscopic redshift of $z=1.150$ for this object;
this places it in the lowest quartile of the submm galaxy redshift distribution.

{\it CUDSS 14.18}.  For this object
our 8.0~$\mu$m search selects the same candidate as Webb \etal found at $K$.
This is identical to the I-selected candidate from Lilly \etal (1999) as well.
Within the positional uncertainties, all three positions 
(IRAC+$K$-selected+I-selected) are also coincident with 
the ISO 7 and 15~$\mu$m detections.  
The 8.0~$\mu$m object is also a bright 24~$\mu$m source.  Finally, the best IRAC position 
is coincident with that observed at Keck by Chapman \etal (2005), who found 
a spectroscopic redshift of $z=0.661$ for this object.  This is a
surprisingly low redshift compared to most SCUBA sources.

{\it CUDSS 14.19}. For this object our 8.0~$\mu$m search selects 
a different best candidate than at $K$.  The IRAC candidate position 
lies just 2\arcsec\ from the radio position 
and is coincident with significant 24~$\mu$m emission.  

\subsubsection{Secure Identifications for Sources without Radio Identifications }

We identify what we believe to be secure infrared identifications 
for two objects that lack radio counterparts, 14.17 and 14.23.

{\it CUDSS 14.17}. Our 8.0~$\mu$m search selects a single best candidate and
disagrees with the $K$-selected candidate, which is not apparent in the mid-infrared mosaics.  
The 8.0~$\mu$m candidate falls just 0\farcs8 from the ISO detection and is detected by MIPS
at 24~$\mu$m.  Thus despite the large offset (9\farcs5) 
this identification appears secure. 

{\it CUDSS 14.23}. Our 8.0~$\mu$m search selects two candidates.
One (B) is fainter, blue in $[3.6]-[4.5]$, and offset 6\farcs6 to the 
NE of the SCUBA position.  The other, brighter source is red in 
$[3.6]-[4.5]$ and $K-[3.6]$ and is a strong source of 24~$\mu$m emission.
It is offset only 2\arcsec\ from the SCUBA position.
For these reasons we prefer candidate A.
It is the same as the $K$-selected candidate from Webb \etal 

\subsubsection{Possible Identifications}

We identify plausible infrared counterparts to ten sources, none of which
has a radio identification (but see discussion of 14.8 and 14.20 below).

{\it CUDSS 14.2}. Although we detect the Webb \etal (2003) candidate (14.2A in 
Table~\ref{8micids}) in all IRAC bands, we also find a closer, fainter IRAC 
object (B) only 2\farcs1 from the reported SCUBA position.  
However, candidate A is a source of MIPS 24~$\mu$m emission and is
much brighter at 8.0~$\mu$m.  For these reasons we adopt A as the more likely
counterpart to 14.2.   The Lilly \etal (1999) I-selected counterpart, 
which is distinct from both the $K$- and IRAC-selected positions, 
is not detected in the IRAC mosaic at 8.0~$\mu$m.

{\it CUDSS 14.4}. At 8.0~$\mu$m there are three possible counterparts. 
Of these, C is approximately coincident with the Webb \etal candidate.  
Object C is also the faintest of the three candidates at 8.0~$\mu$m 
and shows no significant 24~$\mu$m emission.
It is blue in $K - [3.6]$, 
suggesting this object resides at $z < 0.6$.  
Thus C seems unlikely to be the correct counterpart.
Object A presents a red $[3.6] - [4.5] > 0$ color,
but the $K - [3.6]$ upper limit is not sufficiently sensitive to
constrain the redshift.
Object A is a MIPS 24~$\mu$m source.  Object B has no counterpart in the 24~$\mu$m 
catalog, but this is most likely because it is blended with an even brighter 
object to the NW.
However object B presents a blue $[3.6] - [4.5]$ color, suggesting it lies
at $z < 1.3$.  
Thus candidate A seems the most likely SCUBA counterpart among the 
IRAC sources,
despite its relatively large separation
(8\farcs7) from the SCUBA position.

{\it CUDSS 14.7}. There are two possible 8.0~$\mu$m counterparts within the 10\arcsec\ 
search radius.  One of them (A, the closer and fainter of the two) coincides with 
the Webb \etal (2003) $K$-selected candidate.  
Although both are detected in all IRAC bands, only the southern source 
(B, the one more distant from the reported SCUBA position) is unambiguously detected by MIPS.  
The southern source 
exhibits multiple components
in our deep R-band image.  It also has a redder $[5.8]-[8.0]$ and $[8.0]-[24.0]$ colors 
and is by far the brighter source at 8.0~$\mu$m.  For these reasons we prefer candidate B.
However, the SCUBA source is extended toward both IRAC objects.
In addition, both IRAC candidates have red $K$-to-4.5~$\mu$m colors that
suggest they both reside at $z>1.3$.  Thus while we prefer B we cannot rule out 
the possibility that both objects may contribute to the submillimeter flux.

{\it CUDSS 14.8}. Our 8.0~$\mu$m search selects two candidates (A and B).
There is emission at the location of the $K$-selected candidate (B), but
it is present at less than 5$\sigma$ significance.  
Candidate A is roughly 6\farcs5 west of the best SCUBA position.  
There is 24~$\mu$m emission at this position but not
at the position of candidate B.

Interestingly, the $K$-selected candidate may be a source of radio emission
(Chapman \etal 2005), and a spectroscopic redshift $z=2.128$ has been 
obtained for it. The radio detection is regarded as somewhat marginal however 
($\approx3\sigma$, S. Chapman 2005, private communication), and so
the association of candidate B with the SCUBA source is not
established with certainty.
Object B is detected at 3.6 and 4.5~$\mu$m with flux densities of 
5.4 and 5.6~$\mu$Jy, respectively,
but it exhibits no significant emission at 5.8, 8, or 24~$\mu$m.
Because B has a blue mid-infrared SED and only A is significantly 
detected 8.0~$\mu$m, we prefer A as the counterpart.
This object has red $K$-to-4.5~$\mu$m colors that place it at $z>1.3$, meaning it is possible
that it could reside at the same redshift as candidate B as part of
an interacting system.

{\it CUDSS 14.10}. Our 8.0~$\mu$m search selects a different best candidate (A) than
Webb \etal (B), 9\farcs8 SW of the best SCUBA position.  
Candidate A shows $K-[3.6] = 1.5$ and $[3.6]-[4.5]=-0.04$, suggesting it lies in 
the range $0.6 < z < 1.3$.  Candidate B, the $K$-selected object, 
though closer to the SCUBA position (4\farcs4 away),
is almost 1 magnitude fainter at K, is blue in $[3.6]-[4.5]$, 
and exhibits no significant emission at 5.8, 8, or 24~$\mu$m. 
Hence A is the more likely counterpart.

{\it CUDSS 14.12}. Our 8.0~$\mu$m diagnostic selects two candidates, 
neither of which is coincident with the $K$-selected candidate. 
The $K$-selected candidate is not detected in the 8.0~$\mu$m mosaic.
Candidate A lies 5\farcs5 W of the SCUBA coordinates and 
exhibits significant 24~$\mu$m emission. 
Neither A nor B is a source of significant 5.8~$\mu$m emission.
Although B is brighter at 8.0~$\mu$m than A, it exhibits no significant 24~$\mu$m 
emission.  Hence A is the more 
likely candidate, although the identification is by no means secure.

{\it CUDSS 14.14}. At 8.0~$\mu$m there are two candidates.  Neither is coincident 
with the $K$-selected candidate.  Significant MIPS 24~$\mu$m emission 
is associated only with candidate B.  Likewise only this candidate has significant 
detections in all four IRAC bands.  Furthermore, only candidate B is red in $[3.6]-[4.5]$.
Candidate B is our preferred counterpart on the basis of its relative 
brightness, the IRAC colors, and the presence 
of 24~$\mu$m emission at its position.

\begin{figure}
\epsscale{.9}
\figurenum{1}
\vskip -1.3in
\plotone{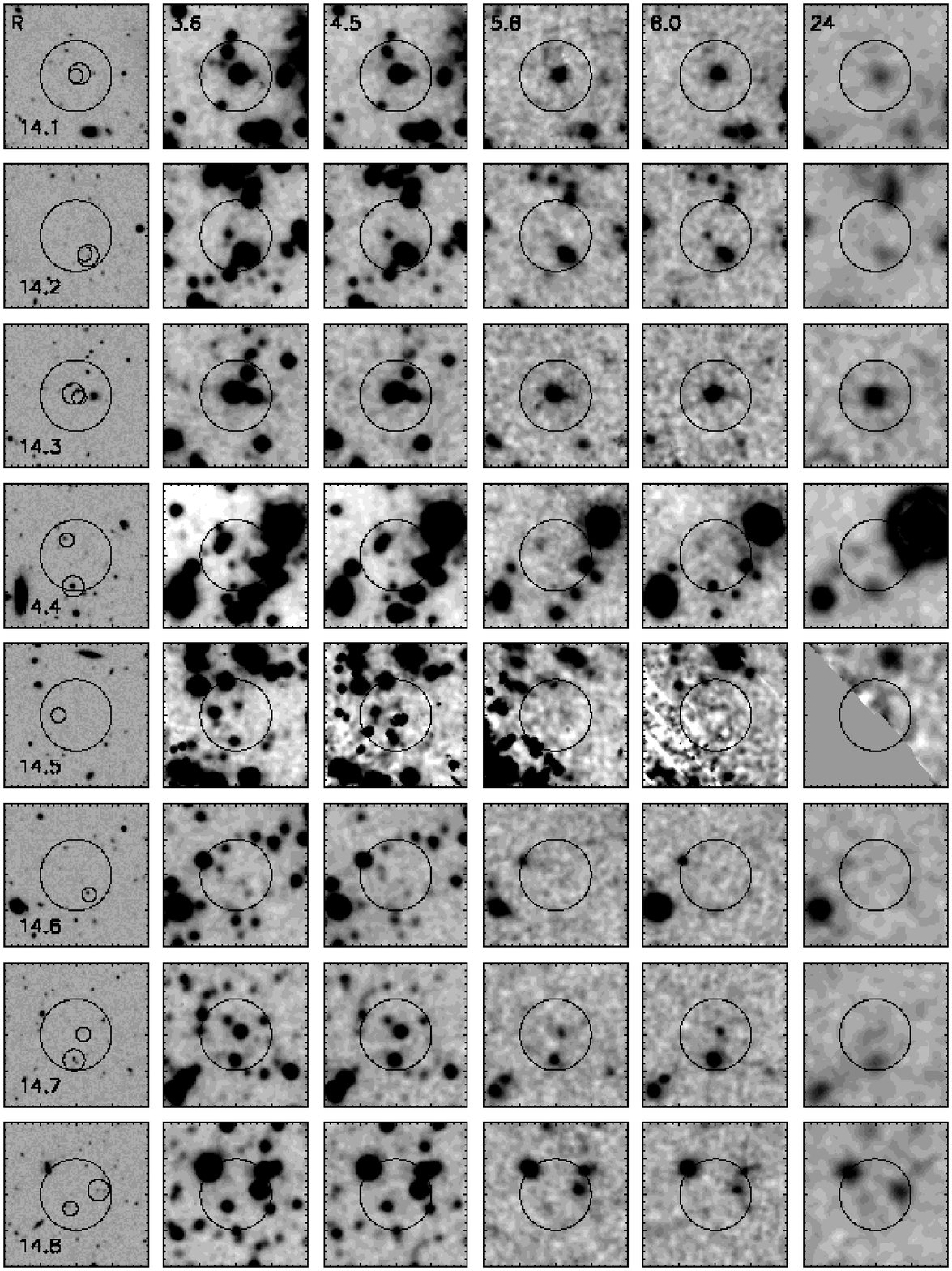}
\caption{R+IRAC+MIPS 24~$\mu$m mosaics at the positions of SCUBA sources 1-8 in the
14$^{\rm h}$ field. Each image is 40\arcsec\ across and centered at
either the SCUBA position (most sources) or the 1.4 GHz radio
position where available (sources 14.1, 14.3, 14.9, 14.13, 14.18, and 14.19).
Circles of 10\arcsec\ radius have been drawn to indicate the area searched
for IRAC counterparts. Proceeding from left to right, the columns
contain images taken at R, 3.6, 4.5, 5.8, 8.0, and 24~$\mu$m.
Circles of diameter 2\arcsec\ and 3\arcsec\ in the R-band images
respectively indicate the locations of the K-band and 8.0~$\mu$m counterparts.
  \label{fig:montage}}
\end{figure}

{\it CUDSS 14.15}.  There are two possible 8.0~$\mu$m counterparts.
One of these (A, the brightest) is coincident with the best $K$-selected candidate 
and is a MIPS 24~$\mu$m source.  Candidate B, however, exhibits comparable 24~$\mu$m 
emission, has a red $[3.6]-[4.5]$ color, and has an SED that peaks at 5.8~$\mu$m, suggestive
of a galaxy at $z\sim2.5$.  Although A is brighter at 8.0~$\mu$m,
B is a more likely candidate because of these SED features.  Perhaps because the $K$-band
coverage at this location is relatively shallow, only a $K$ upper limit can
be determined for candidate B.  The R-band source seen in Figure~\ref{fig:montage}
close to the apparent position of candidate B is not the visible-wavelength counterpart.
Only upper limits could be established in the bands blueward of I for this object.

{\it CUDSS 14.20}. The 8.0~$\mu$m search selects a different 
best candidate than Webb \etal.  
The IRAC candidate is not detected in our deep R-band image 
or the Webb \etal I-band image but is detected with $K = 22.26\pm0.06$ mag.
It therefore has red $K$-to-4.5~$\mu$m colors that place it at $z>1.3$.
There is no significant MIPS 24~$\mu$m source within 10\arcsec\ of the SCUBA position.
Acquiring a visible-wavelength spectrum of 
this extremely optically-faint object would be very challenging. 
As is the case for 14.8, we select a different position than preferred
by Chapman \etal (2005), who obtained a spectroscopic redshift of $z=2.128$
for the $K$-selected candidate, where they detect 1.4 GHz emission.  
However, the radio identification, which is made at roughly
4$\sigma$ significance (S. Chapman 2005, private communication) 
may be incorrect.

{\it CUDSS 14.22}.  The 8.0~$\mu$m search finds two candidates.
One of these (A) is coincident with the the $K$-selected candidate to
the NW.  Even though both are red in $K-[3.6]$, 
the fact that they are both blue in $[3.6]-[4.5]$ suggests that they
lie in a relatively nearby redshift range $0.6 < z < 1.3$ and 
that we may have not detected the true counterpart of this object.
However, only the southernmost candidate is a strong 24~$\mu$m source,
so we regard it as the more plausible IRAC counterpart.

For the sake of completeness, we examined the relatively faint 24~$\mu$m source
apparent in Figure~\ref{fig:montage} 10\farcs3
NNW of the SCUBA position, just outside the search radius.
It is blue throughout the IRAC bands and is a much weaker 24~$\mu$m source
(232~$\mu$Jy) than candidate B.  It is therefore likely to be a low-redshift
object and not associated with the SCUBA source.

\subsubsection{Failed Identifications}

{\it CUDSS 14.6}. 
Despite the excellent IRAC coverage at the location of CUDSS 14.6 (a total of 
218 exposures), no sources in the IRAC 8.0$\mu$m mosaic satisfy our selection 
criteria as a valid counterpart to CUDSS 14.6.  
Within 10\arcsec\ of the reported SCUBA position, and specifically at
the position identified by Webb \etal (2003), there is no significant
5.8, 8.0, or 24~$\mu$m emission.  Moreover, the {\sl K} to 4.5~$\mu$m colors
of the {\sl K}-selected counterpart are blue, suggesting that object lies nearby.

There is a faint source in the 8.0~$\mu$m mosaic within the 10\arcsec\ search 
radius, 8\farcs5 to the NE of the SCUBA position.  However, its low 
flux density (5.5~$\mu$Jy) puts it below the 5$\sigma$ sensitivity of the 
8.0~$\mu$m mosaic.  Like the {\sl K}-selected counterpart it also has a blue 
$K-[3.6]$ color, implying that it resides relatively nearby.  The object appears in
three of the IRAC mosaics (it is not apparent at 5.8~$\mu$m) and is therefore
unquestionably real, but it is too faint to satisfy our selection criteria.

A significant 8.0~$\mu$m detection (18~$\mu$Jy) is located just
outside the search area, 10\farcs1 to the NE of the SCUBA position.
It is red in $K-[3.6]$ but exhibits a blue $[3.6]-[4.5]$ color, suggesting
it lies at an intermediate ($0.6 < z < 1.3$) redshift and is therefore
unlikely to be the true counterpart.

Both of the 8.0~$\mu$m sources mentioned above are as likely as not to be
due to chance superposition.  In combination with the color information
this strongly suggests that no plausible counterpart to the SCUBA source
is present in the IRAC mosaics.

{\it CUDSS 14.11}. At 8.0~$\mu$m there are two indistinguishable best candidates, one of which
(B) is coincident with the $K$-selected candidate.  Candidate A is 
closer to the SCUBA position, only 2\farcs6 away.  Neither shows any 
significant 24~$\mu$m emission, and both exhibit the morphologies of 
elliptical galaxies in our deep R-band image.
Furthermore, the blue infrared colors ($K-[3.6] = -0.53,-0.60$, and $[3.6]-[4.5]=-0.48,-0.69$)
and {\sl R}-band morphologies are consistent with both objects being elliptical 
galaxies at $z<0.6$.
There is no significant 24~$\mu$m emission within 10\arcsec\ of the SCUBA position.  
For all these reasons, we conclude that we have not detected the true counterpart 
to this SCUBA source.  A speculative possibility is that the 850~$\mu$m
emission detected by SCUBA is lensed by the foreground, low-redshift objects.
The closest 24~$\mu$m source (176~$\mu$Jy) lies 15\farcs7 NE of the SCUBA position
and has blue colors throughout the IRAC bands.

{\it CUDSS 14.21}.  There are no significant 8.0~$\mu$m sources
within 10\arcsec\ of the SCUBA position of this source.
We therefore attempt no identification.  There is significant 3.6 
and 4.5~$\mu$m emission at the location of the $K$-selected counterpart.
This object is red in $K-[3.6]$ but blue in $[3.6]-[4.5]$, suggesting
that it may be at $0.6 < z < 1.3$.
At 5.8, 8.0, and 24.0~$\mu$m only upper limits are available.
We regard CUDSS 14.21 as unidentified at 8~$\mu$m.

Assuming that these sources are not highly dust-obscured galaxies at
such extremely high redshift ($z>6$) that their obscured rest-visible emission 
falls in the observed mid-infrared, we suggest three possible causes for the 
lack of plausible 8~$\mu$m counterparts.  In the case of CUDSS 14.11, 
it may be that the two foreground elliptical galaxies are screening 
a high-redshift SCUBA source from view at visible-mid-infrared wavelengths.  
A second alternative
is that the submillimeter flux densities may include contributions from multiple
weak components that lie within the SCUBA beam but not precisely at 
the reported SCUBA position.  The SCUBA maps are known to be highly
confused below 2~mJy (Knudsen \etal 2006, and references therein).  
Weak submillimeter sources that combine to yield an apparent SCUBA detection
might well be individually too faint to detect even in the deep IRAC mosaic.
Third, one of the SCUBA sources might be spurious, which 
would not be unreasonable given that the sample contains so many 
roughly 3$\sigma$ sources.  CUDSS 14.6, 14.11, and 14.21 are
detected in the 850~$\mu$m maps at 4.2, 3.5, and 3.0$\sigma$ significance,
respectively.

\subsubsection{Objects with Insufficient Coverage}

Three sources lie in a portion of the IRAC 8.0~$\mu$m survey region for which
fewer than 10 exposures were acquired.  For these sources (14.5, 14.9, and 14.16)
the depth of coverage was insufficient to reliably detect an infrared counterpart.
%
%
%
%

\subsection{Reliability of IRAC Detections}
\label{sec:detection}

Excluding `dubious' associations, potential counterparts for 17 of 20 SCUBA galaxies 
having deep 8.0~$\mu$m coverage in the 14$^{\rm h}$ field have been identified (Table~2).  
Of these, the seven secure identifications have a mean offset from the SCUBA positions
of 4.4" and the ten possible identifications have a mean offset of 7.7".
If one uses the available 1.4 GHz positions in preference to the SCUBA positions
only the mean offset for the set of secure identifications changes (because no
SCUBA sources with `possible' identifications have radio positions), to 2\farcs4.
The mean offsets in right ascension and declination for the 17
counterparts are 2\farcs5 and 0\farcs9 (SCUBA - 8.0~$\mu$m), respectively.
All these offsets are smaller than the SCUBA positional uncertainties.

Of course, a small number of the 8.0~$\mu$m-selected counterparts are chance associations 
and not physically associated with the corresponding SCUBA source.  In the following
we address this issue in two ways.  First, we derive an estimate of the number of such 
chance associations using a statistical analysis.  Second, we compare the infrared-submm
colors of our counterparts to those that have been reliably identified in a deep
IRAC survey of the Lockman Hole region by Egami \etal (2004).

We first employ a variant of the statistical formalism used by Lilly \etal (1999)
and others to assess the significance of possible matches.
Specifically, we estimate a statistic, usually denoted $P$,
that a candidate IRAC source lies within the search distance (10\arcsec)
of a SCUBA source to which it is unrelated.  The statistic is
$P = 1 - {\rm exp}(-\pi n r^2),$ where $r=10$ and $n$ is the surface density
of IRAC sources at least as bright
as the candidate.  Thus lower values of $P$ suggest a higher likelihood
that the association of the IRAC and SCUBA sources is significant.
Unlike Lilly \etal (1999) we treat all sources within the 10\arcsec\
search radius equally with respect to their offset from the SCUBA
position.  In doing so we avoid unduly biasing our selection toward
sources at small offsets from the uncertain SCUBA positions.

We use Table~1 of Webb \etal (2003) for the SCUBA sources' coordinates,
except for CUDSS 14.1, 14.3, 14.13, 14.18, and 14.19.  The first four
of these objects have VLA positions (Eales \etal 2000) with lower uncertainties
(roughly 2\arcsec).  Similarly, we adopt the radio position reported for 14.19 by
Webb \etal (2003), although it is offset from the nominal SCUBA position by 8\farcs5.
We found no significant differences in our lists of most-likely matches when 
we repeated the analysis using the less-certain SCUBA positions.

In order to interpret the $P$-values, we performed a Monte Carlo analysis
in which randomly-generated catalogs of IRAC sources were searched for
counterparts in a manner identical to that employed for the real sources.
The random catalogs were constructed to present brightness distributions and source
densities identical to those of the real catalog.
The $P$-distributions for the random and real samples are compared in
Figure~\ref{fig:pdist}.  We calculate the probability $P^\prime$
of finding a random or unassociated source as bright as the observed
candidate as $P^\prime = \alpha P$, where
$\alpha$ is the ratio of the number of random sources with smaller
values of $P$ than the candidate to the total number of random sources
generated in the Monte Carlo calculation.
Although the $P^\prime$ statistics are not meaningful for individual sources, 
the sum of the $P^\prime$ statistics tabulated in Table~\ref{8micids}
is an estimate of the number of spurious identifications.
We cannot say which {\it particular}
identifications are wrong, but the sums for the seven secure
and the ten possible identifications are 0.61 and 2.61, respectively, 
implying that the
bulk of the errors apply to the latter subset, as one would expect.  
For the sample as a whole, the sum is 3.22, suggesting that
about three of our 17 identifications are incorrect.  
This estimate (three false identifications) is a pessimistic one,
because the probability formalism from which it is derived does not account for
our use of color information to winnow lists of multiple candidates 
down to single most-likely counterparts.  Three is in fact the 
maximum number of likely false sources, because we have taken into account 
the $K$ to mid-infrared colors as well as the presence of 24~$\mu$m MIPS 
emission as identification criteria.

When the estimate for the number of spurious sources is subtracted, our
identification rate is (conservatively) 14/20, or 70\%.  
This is comparable to or better than rates described elsewhere.  
Borys \etal (2004) recovered 11 of 19 (57\%) SCUBA source in the HDF-N.
Pope \etal (2005) reported a 72\% recovery rate for SCUBA 
sources in the GOODS-N field, although this figure doesn't account for possible 
erroneous associations.  Like Pope \etal (2005) we recovered essentially all the 
SCUBA sources with secure radio positions (but see the discussions of 14.8 and 
14.20 above).  More straightforward comparisons can be made with Lilly \etal (1999)
and Webb \etal (2003), because the $P^\prime$ statistics they provide 
permit one to estimate the number of incorrect associations in a manner
identical to that applied to our own sample.  Lilly \etal (1999) identified
counterparts for eight 
of the 12 submillimeter sources they observed at visible wavelengths.
The sum of the resulting $P^\prime$ statistics was $\sim1$, suggesting
a total identification rate of 7/12 or 58\%.
Likewise the Webb \etal (2003) $K$-band counterparts for which the $P^\prime$ statistics
sum to $\sim6$, implying a total identification rate of 17/23 or 74\%
(but see below).

Egami \etal (2004) presented mid-infrared through submillimeter flux densities measured 
by Spitzer for SCUBA sources in a $5^\prime\times5^\prime$ area of the Lockman Hole East.
Because these sources have radio counterparts from Ivison \etal (2002) their positions
are well-localized, and the mid-infrared identifications are secure.
For this reason, the Egami \etal (2004) submillimeter sample provides a useful
basis with which to assess our own larger sample of infrared-selected counterparts.

Figure~\ref{fig:irac_colors} shows 
the distributions of submillimeter galaxies in the IRAC $[3.6]-[4.5]$ 
versus $[5.8]-[8.0]$ color-color space.  
The portion of this space occupied by the Lockman Hole sources and the secure
identifications of the CUDSS 14$^{\rm h}$ field sources are approximately
coextensive; they define a locus in which other counterparts sources may
be expected to fall.  This is indeed the case for the `possible' sources
taken as a group -- the distribution of these objects occupies a very similar
region of this color-color space overall as do the secure sources.  

In general, when the counterpart is bright, the $K$- and 8.0~$\mu$m-selected 
counterparts are identical.
However, the mid-infrared technique selects different counterparts than Webb \etal (2003) 
in 10 cases (CUDSS 14.4, 14.7, 14.8, 14.10, 14.12, 14.14, 14.15, 14.19, 14.20, and 14.22).  
The $K$- and 8.0~$\mu$m-selected techniques disagree for three additional cases 
(CUDSS 14.6, 14.11, and 14.21) for which the mid-infrared technique identifies
no plausible counterparts.  CUDSS 14.11 is a nearby elliptical 
galaxy and is therefore unlikely to be the source of the submillimeter emission.  
CUDSS 14.6 and 14.21 have no plausible counterparts in the 8.0~$\mu$m mosaics.  
The $K$- and 8.0~$\mu$m-selected counterparts therefore disagree in a total of 13 cases
(slightly more than half the total).  With the exception of 
the counterpart to CUDSS 14.7A (which as noted above may indeed be a contributor
to the submillimeter emission), the distribution of the $K$-selected sources is
clearly blue in both colors compared to the other objects.  In particular
it lies close to the region occupied by a passively evolving elliptical galaxy. 
Only six $K$-selected counterparts are plotted because the remainder are undetected
at both 5.8 and 8.0~$\mu$m, however the $[3.6]-[4.5]$ colors of all these other sources
tend to be similarly blue in $[3.6]-[4.5]$ and not like the distributions of the
secure identifications (Table~\ref{8micids}).  
We conclude that the 8.0~$\mu$m-selected counterparts are therefore
more likely to be the correct counterparts in these discrepant cases. 

Figure~\ref{fig:mips_scuba} shows the same populations
but in $[24]-[850]$ versus $[8.0]-[24]$ color-color space.  Again, the distributions
of the Lockman Hole counterparts and the secure sources from this work are
similar. CUDSS 14.13 and 14.18 are understandable exceptions on account of
their relatively low redshifts.  And as before, the `possible' counterparts
occupy a region of this color-color space that is reassuringly similar 
to that of the secure sources.  That is not the case for the discrepant
$K$-selected counterparts.  These objects lie at the outskirts of 
the secure counterparts' distribution.
Based on the Lockman Hole sources and the secure 8.0~$\mu$m-selected
sources (not including CUDSS 14.13 and 14.18), one would expect the typical 
$[24]-[850]$ color of a SCUBA galaxy to be between 3 and 4 magnitudes.
Not one of the discrepant $K$-selected counterparts lies in this range, and
indeed only one (14.15A) is even detected at 24~$\mu$m.  

Analogous trends are apparent in $[3.6]-[8.0]$ versus $[8.0]-[24]$ color-color space
shown in Figure~\ref{fig:mips_irac}.  
The secure 8.0~$\mu$m-selected sources occupy the
same region of this color-color space as the Lockman Hole sources, and that
region is spanned by the possible 8.0~$\mu$m identifications.  
The low-redshift objects are not outliers in this distribution.  As before, the
discrepant $K$-selected counterparts tend to be blue in $[3.6]-[8.0]$ and
with the exception of 14.7A do not overlap with the securely identified counterparts; 
half of these discrepant sources are so blue they are not even detected at 8 or 24~$\mu$m.

\section{Discussion}
\label{sec:discussion}

\subsection{SEDs of SCUBA Sources Detected by IRAC}
\label{sec:seds}

Figure~\ref{fig:sedplot} shows the SEDs of our most-likely IRAC counterparts 
to the SCUBA sources in this field, showing the diversity of spectral shapes
present within the sample.  The galaxies divide roughly into two categories.  
The majority (12 objects) show evidence in the IRAC bands
for the 1.6~$\mu$m bump arising from the H$^-$ opacity minimum.  
In other words, these objects appear to be powered primarily by stars
and not by a compact object.
CUDSS~14.18 is among these sources, in agreement with Chapman \etal (2005).  

Four objects in the sample (sources 14.3, 14.7, 14.13, and 14.19) show the
monotonically rising SEDs indicative of AGN-dominated emission.
These sources fulfill all three of the IRAC-color AGN
selection criteria proposed by \citet{lacy04}, \citet{stern05}, and \citet{hatz05}.
Source 14.4 has an unusual SED, very faint in the three shortest-wavelength
IRAC bands compared to the optical and 8 and 24~$\mu$m.  This source
also has a monotonic rise through the IRAC bands, so we consider it
a tentative AGN candidate.

Simple color criteria developed from shallow IRAC surveys like those cited
above are not adequate to separate AGN from starburst galaxies at high redshift
however.  We therefore chose to perform a simple SED fitting procedure to
determine whether a galaxy is AGN-dominated, in exactly the same way 
as \citet{bar06} did for other sources in the Extended Groth Strip.

Specifically, we fit power laws ($f_{\nu}\propto {\nu}^{\alpha}$) to IRAC flux 
densities as an empirical way of characterizing their SEDs.  Seven sources have acceptable
power-law fits.  Source 14.14 has a blue power-law ($\alpha \approx +0.2 $),
source 14.18 has $\alpha \approx 0$, and five others 14.3, 14.7, 14.12, 14.13, and 14.19)
have good fits to red power-laws.  Source 14.4 does not fit a power-law well 
because of a very steep rise in the SED at 8.0~$\mu$m.

AGNs thus selected on the basis of the IRAC-MIPS SEDs show good agreement
with AGNs discovered on the basis of their X-ray emission.  Four objects 
with secure mid-infrared power-law AGN signatures are X-ray sources
(14.3, 14.7, and 14.19 are Nandra \etal 2005 objects c111, c113, and c128 
respectively; 14.13 is source c72 in the list of Nandra \etal and source 23 in 
Waskett \etal 2003).  CUDSS 14.12 and 14.4 (the tentative AGN candidate), 
show no X-ray emission.

Our classification of 14.13 as AGN-dominated agrees with Higdon \etal (2004) and
Le Floc'h \etal (2006, in preparation), although Chapman \etal (2005) call it a starburst.
Given the small numbers involved, our finding that between five and six 
members of the sample are AGN-dominated is not inconsistent with the assertion
by Alexander \etal (2004) that ``at most 20\%" of the luminosity of submillimeter
galaxies arises from AGN.  They found that bright ($F_{850\mu m} > 5$ mJy) 
SCUBA sources host AGNs at twice this rate (38\%), although star formation tends
to dominate the energy output.  

In summary, the number of SCUBA counterparts apparently dominated by an AGN 
in the mid-infrared could be as few as five (CUDSS 14.3, 14.7, 14.12, 14.13, and 14.19) 
or as many as six (if 14.4 is included).  There could be others not detectable
using our approach: \citet{bar06} found that not all (X--ray selected) AGN show 
clear mid-infrared signatures.

Figure~\ref{fig:fluxhist} shows the 850~$\mu$m flux density distribution of the 23
SCUBA galaxies.  Our 17 counterparts span the full range occupied by the sample, 
including all of the brightest SCUBA sources and all but one of the faintest.  
There is no evidence that our recovery rate is biased toward either the 
high- or low-brightness SCUBA sources, based on a Kolmogorov-Smirnov (KS) test 
comparing a sample consisting of the retrieved sources with another sample
consisting of all 23 SCUBA galaxies.  Furthermore, 
AGN-dominated objects occupy the middle of the distribution, without any obvious skew
toward either the high- or low-brightness extreme of the distribution.

\subsection{Infrared Colors of SCUBA Galaxies}
\label{ssec:colors}

The infrared color distributions of the IRAC counterparts to the SCUBA galaxies 
are shown in Figure~\ref{fig:histocolor}, and the means and dispersions
are given in Table~\ref{tab:colors}.  Colors at longer wavelengths present 
greater diversity (larger dispersions) both for the SCUBA counterparts and 
for the full sample of all 8$\mu$m-detected galaxies in the field, but
the color distributions are consistently tighter for the counterparts.
In particular, the counterparts do not overlap the blue wings 
of the distributions of the field galaxies.
The SCUBA counterparts are slightly 
redder in the IRAC bands relative to the full sample of all IRAC 
galaxies in the CUDSS 14$^{\rm h}$ field.  Although the difference
in any one color is small, it amounts to nearly 1 
magnitude in $[3.6]-[8.0]$.  This is unlikely to be a selection effect.
The seven counterparts (14.1, 14.3, 14.13, 14.17, 14.18, 
14.19, and 14.20) identified without recourse to any color criteria show no 
significant color differences from the larger sample.

Given the dispersions, the differences in mean colors between SCUBA galaxies
and field galaxies are only marginally significant.  The KS test, however, is
sensitive to the lack of blue galaxies in the counterpart sample.
The KS-derived probabilities of drawing samples of SCUBA counterparts 
that are as different as observed from the sample consisting of all
8~$\mu$m-detected galaxies 
are $6\times10^{-4}$, $10^{-2}$, and $5\times10^{-2}$ for
the $[3.6]-[4.5]$, $[4.5]-[5.8]$, and $[5.8]-[8.0]$ color 
distributions, respectively, if the underlying populations are the same.
In other words, the color distributions are significantly different.

Figure~\ref{fig:mipscolor} shows the [8.0] - [24.0] and {\sl K} - [3.6] 
color distributions for the IRAC-detected galaxies in this field.  
Both distributions exhibit much larger
dispersions than the IRAC-only colors plotted in Figure~\ref{fig:histocolor}.
Despite the fact that in some cases the IRAC counterpart was selected on the basis
of the presence and/or strength of MIPS 24~$\mu$m emission, there are no significant
differences in the mean [8.0] - [24.0] colors or dispersions of the full sample 
and the counterparts.  The {\sl K} - [3.6] color distributions are likewise
indistinguishable (given the wide range of colors exhibited by the sample).
When we compare the color distributions using KS tests we find no evidence
that the samples are different 
with respect to either 
{\sl K} - [3.6] or [8.0] - [24.0] color.

\section{Summary}

Deep IRAC mosaics identify infrared counterparts for 17 of 20 SCUBA galaxies 
in the CUDSS 14$^{\rm h}$ field.  The recovery rate of SCUBA galaxies at 8.0~$\mu$m is 
comparable to or better than that found in other counterpart surveys, 
once the likelihood of mistaken identifications is estimated (three or fewer
for the current work).  For more than half of the sample, the most-likely 
8.0~$\mu$m counterpart is different than the {\sl K}-selected counterpart. 
Multiband IRAC-MIPS SEDs allow one to identify and avoid
unlikely candidates in the SCUBA beam (e.g., elliptical galaxies).
This suggests that the mid-infrared regime
in general (and the IRAC 8.0~$\mu$m band in particular)
offers a more reliable means than ground-based optical or near-infrared
surveys for identifying the true counterparts to the high-redshift, 
dusty starforming galaxies detected in submillimeter surveys.

On the basis of power-law fits to the infrared IRAC-MIPS SEDs 
we infer that five or possibly six of the counterparts are AGN-dominated.
The 8.0~$\mu$m counterparts have redder infrared colors than the general 
population of IRAC-detected galaxies in the field.  Five counterparts
have IRAC-MIPS SEDs that are fit well by red power laws and are likely
to be AGN-dominated.  One more is fit well by a slightly blue power-law
and may also harbor an active nucleus.  Four of these six objects are
detected in X-ray observations, lending further support to the AGN hypothesis.
The remaining counterparts are likely to be dominated by starburst emission.


\acknowledgments

This work is based on observations made with the \sst, which is
operated by the Jet Propulsion Laboratory, California Institute of
Technology under NASA contract 1407. Support for this work was
provided by NASA through Contract Number  1256790 issued by
JPL/Caltech.
We are grateful to the anonymous referee, whose comments and suggestions
greatly improved the clarity and presentation of this work.
IRAF is distributed by the National Optical Astronomy Observatories,
which are operated by the Association of Universities for Research
in Astronomy, Inc., under cooperative agreement with the National
Science Foundation.

Facilities:  \facility{Spitzer(IRAC, MIPS)}.

\clearpage

\clearpage
\plotone{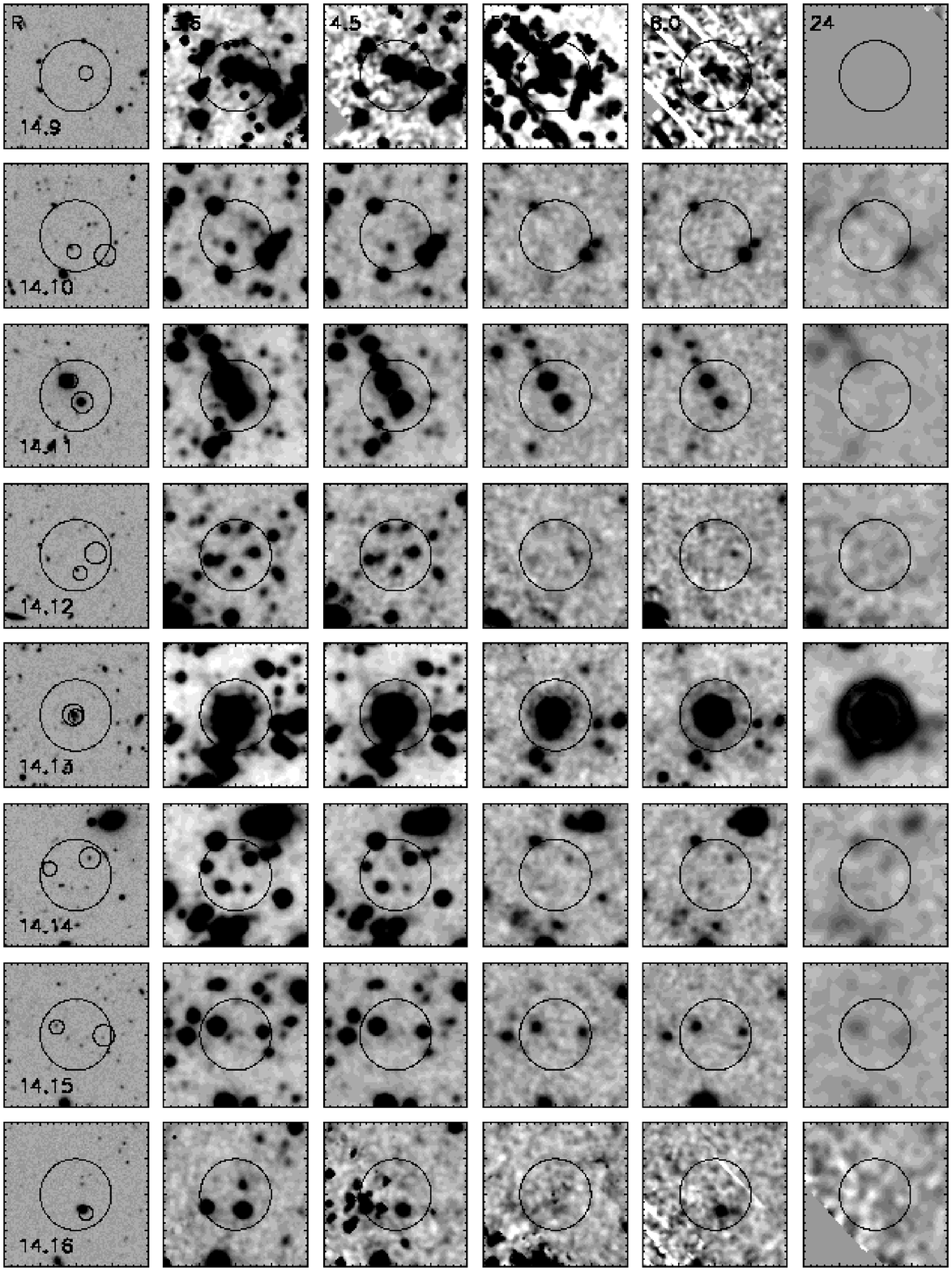}
\centerline{Fig. 1. Continued.}
\clearpage
\plotone{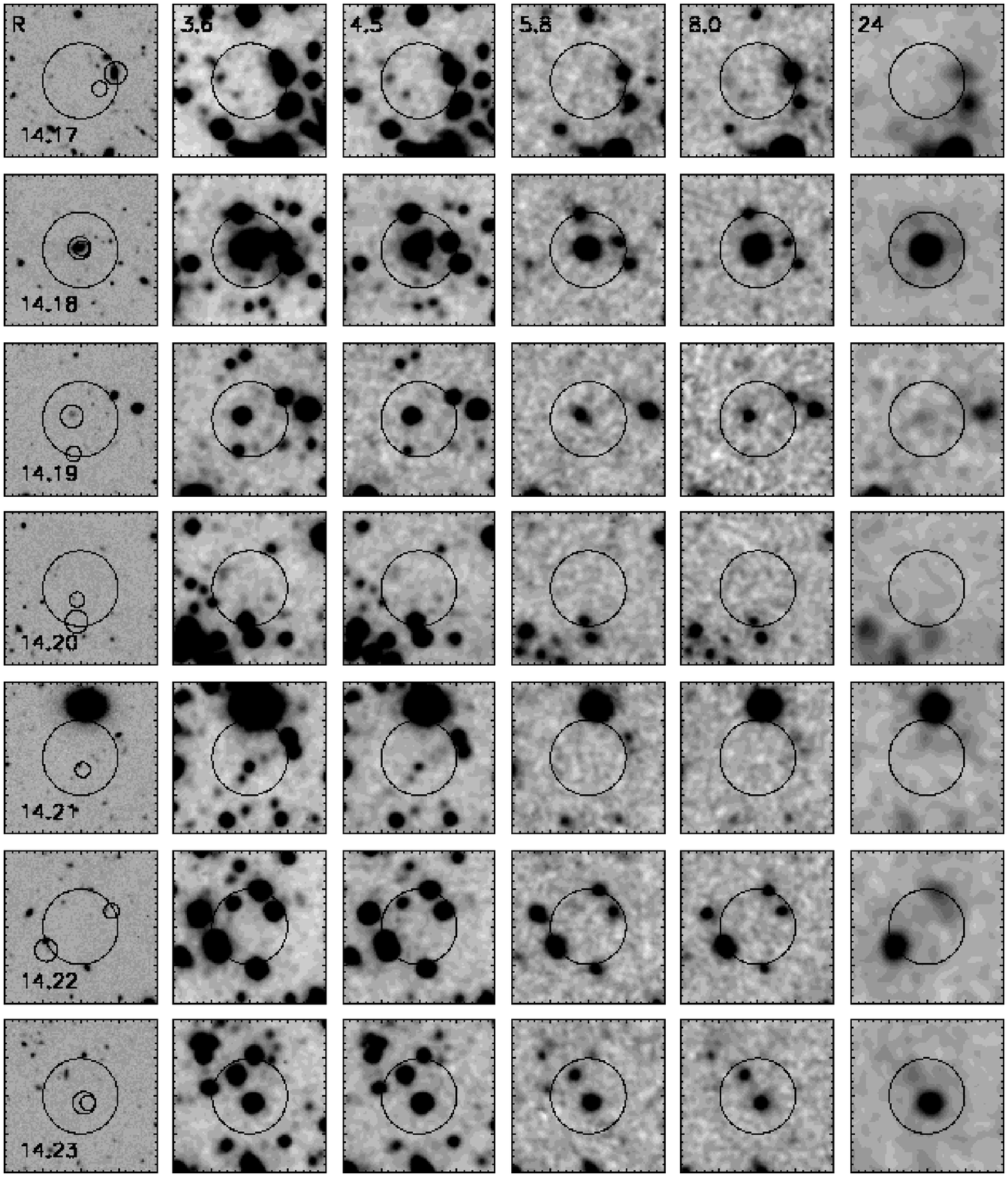}
\centerline{Fig. 1. Continued.}
\clearpage
\addtocounter{figure}{1}

\begin{figure}[!t]
\plotone{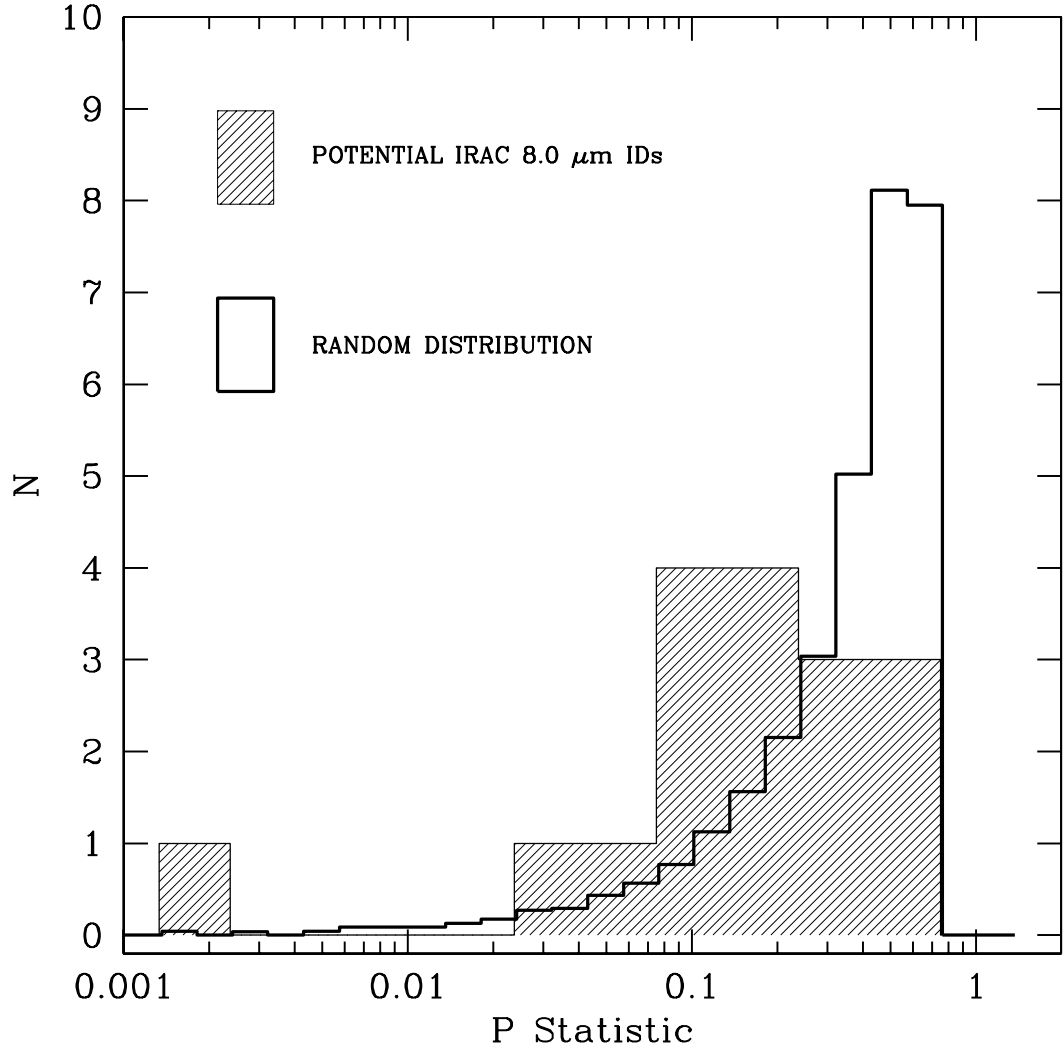}
\caption{The distribution of P-values measured for IRAC 8.0~$\mu$m-selected counterparts
to the SCUBA sources in the Groth Strip.  The shaded histogram indicates the
distribution derived from matches using the real IRAC catalog.  The clear histogram
shows the $P$-distribution derived from matches to sources at random positions. 
\label{fig:pdist} }
\end{figure}  

\begin{figure}[!t]
\plotone{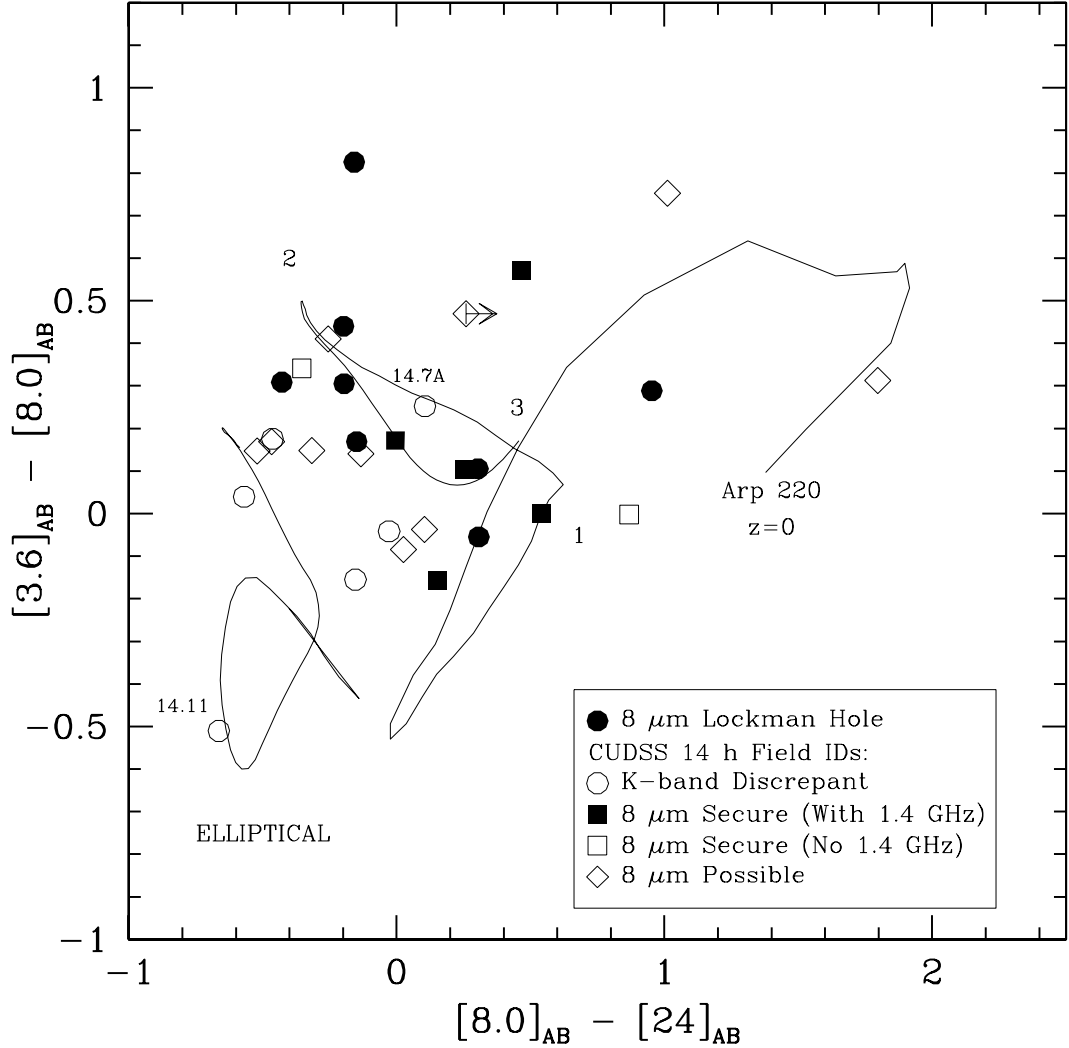}
\caption{The $[3.6] - [4.5]$ versus $[5.8] - [8.0]$
colors for the IRAC and $K$-selected counterparts to submillimeter galaxies.
Solid circles represent radio-detected SCUBA galaxies in the 
Lockman Hole for which secure identifications 
have already been established (Egami \etal 2004).
Secure IRAC identifications for SCUBA galaxies in the CUDSS
14$^{\rm h}$ field are indicated with solid and open squares for
objects respectively with and without 1.4~GHz detections.
Diamonds indicate the `possible' IRAC identifications discussed
in the text.  Open circles represent $K$-selected counterpart identifications
that are different than the most-likely IRAC counterparts.
Also shown are the loci occupied by two nonevolving 
K-corrected galaxy templates as they would be observed in 
the IRAC bands at redshifts from $z=0-3$.  The elliptical 
template exhibits relatively blue infrared colors while those
of the rapidly star-forming Arp~220 are significantly redder 
and approximately coincident with the 8~$\mu$m-selected 
counterparts.
\label{fig:irac_colors} }
\end{figure}

\begin{figure}[!t]
\plotone{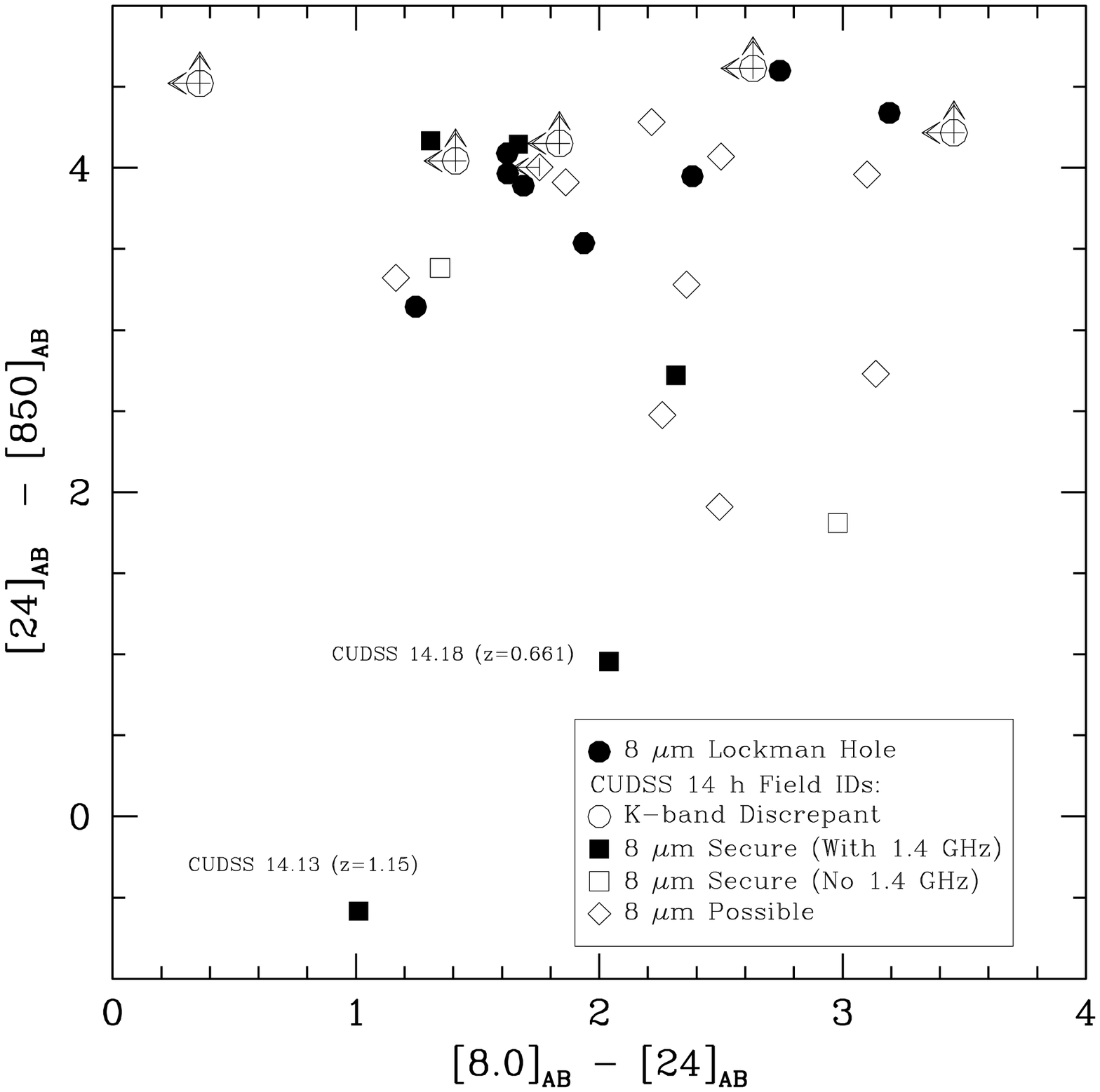}
\caption{The $[24] - [850]$ versus $[8.0] - [24]$
colors for the IRAC counterparts to the submillimeter galaxies. 
Symbols as in Figure~\ref{fig:irac_colors}.
\label{fig:mips_scuba} }
\end{figure}

\begin{figure}[!t]
\plotone{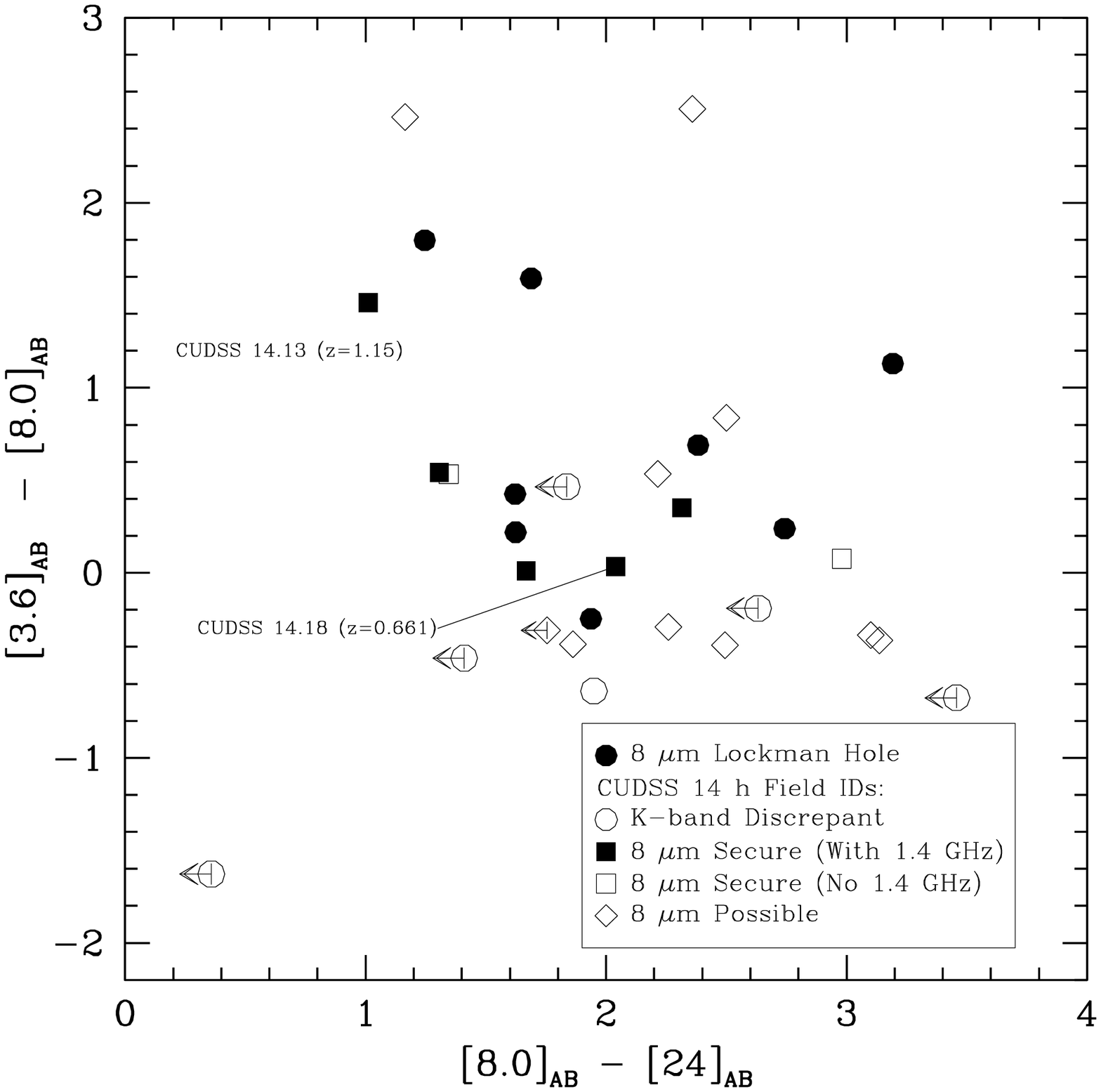}
\caption{The $[3.6] - [8.0]$ versus $[8.0] - [24]$
colors for the IRAC 8.0~$\mu$m counterparts to the submillimeter galaxies. 
Symbols as in Figure~\ref{fig:irac_colors}.
\label{fig:mips_irac} }
\end{figure}

\begin{figure}[t]
\vskip -0.5in
\plotone{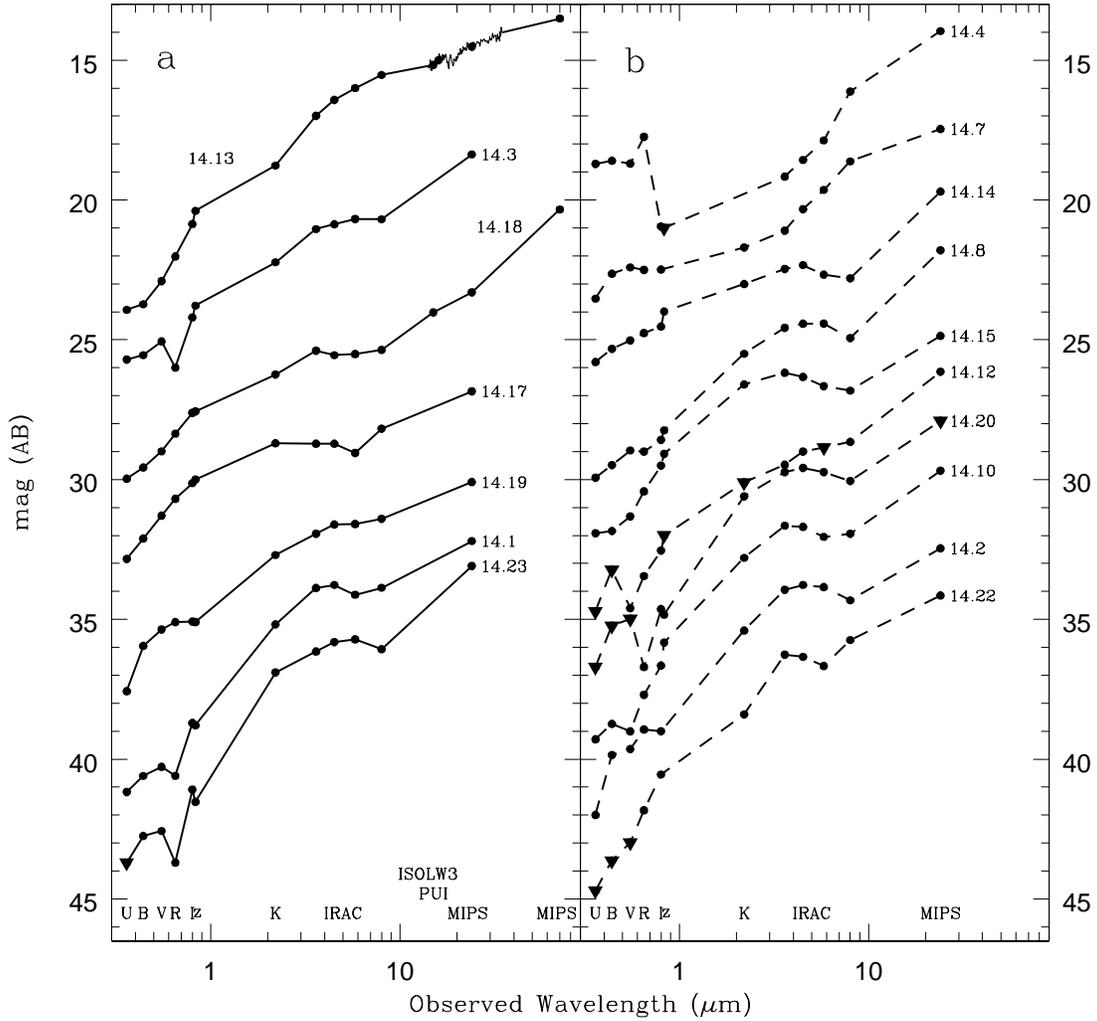}
\caption{Panchromatic {\sl UBVRIzK + IRAC + MIPS 24~$\mu$}m SEDs for the preferred 
IRAC counterparts of the submillimeter galaxies in the 14$^{\rm h}$ field.  
Sources with secure identifications are plotted in panel a with solid lines. 
The `possible' identifications (see Section~\ref{ssec:notes}) are shown in panel b 
with dashed lines.  Triangles indicate upper limits.  For CUDSS 14.13 includes 
the IRS 16~$\mu$m Peakup Imager flux density 3.6 mJy 
as well as the IRS long-wavelength low-resolution spectrograph spectrum
from Higdon \etal (2004), and the ISO 15~$\mu$m magnitude
from Flores \etal (1999).  All other SEDs are offset by arbitrary
amounts for illustrative purposes; the offset amounts are 0, 1, 7, 9, 11 14, and 16 mag
for sources 14.13, 14.3, 14.18, 14.17, 14.19, 14.1, and 14.23, respectively, 
in panel a.  Similarly, in panel b the offsets are 
-4,    -1,   1,     4,    6,     7,     9,      12,   14,   and 17 mag for sources 
14.4, 14.7, 14.14, 14.8, 14.15, 14.12, 14.20, 14.10, 14.2, and 14.22, respectively.
The top two spectra in each panel (14.13 and 14.3 in panel a; 14.4 and 14.7 in panel b) 
are classified as
AGN-dominated based on either their IRAC colors or 
on the monotonic increase in magnitude with wavelength
seen through the IRAC bands.
\label{fig:sedplot} }
\end{figure}  

\begin{figure}[!t]
\plotone{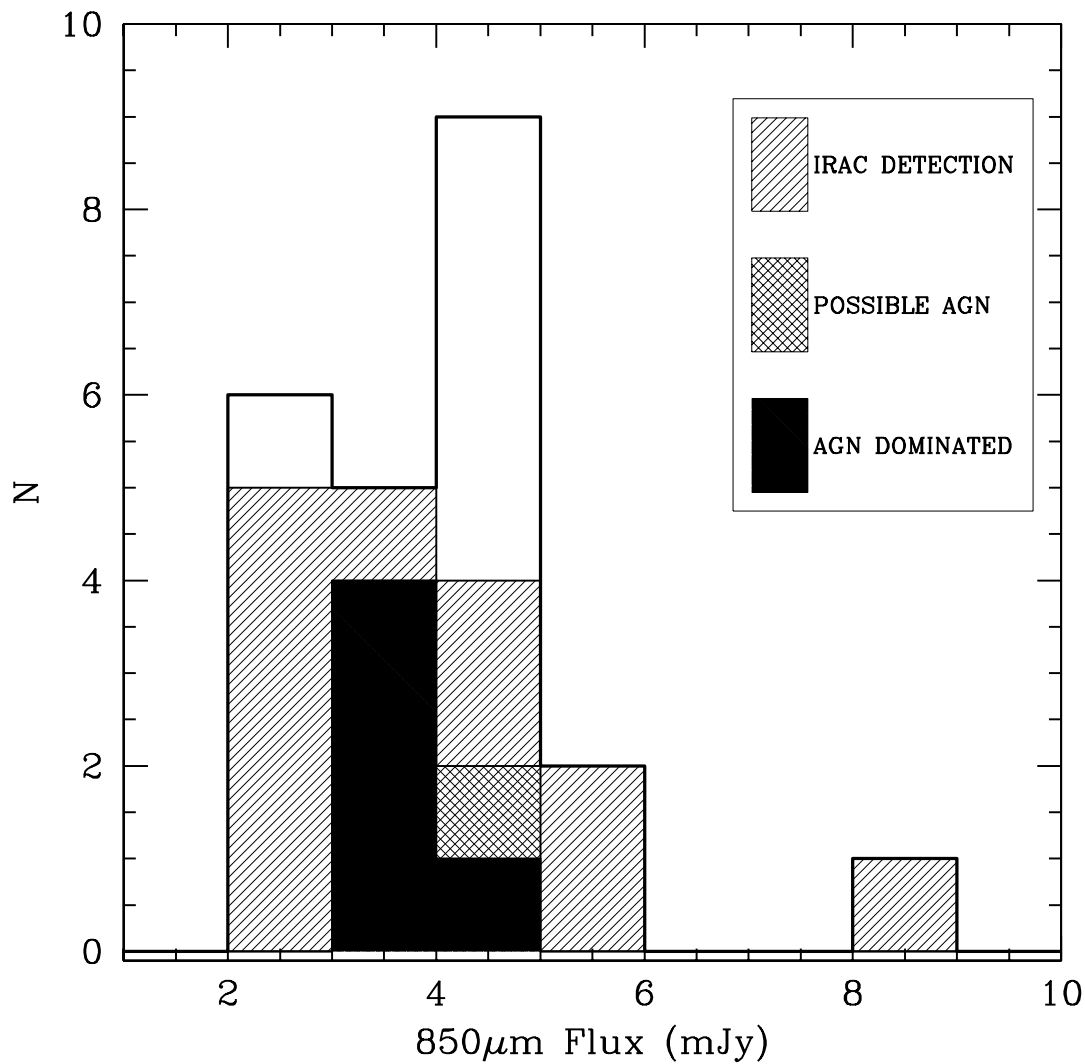}
\caption{Histogram showing the 850~$\mu$m flux density distribution
in mJy for the 23 SCUBA sources in the 14$^{\rm h}$ field.  The shaded portion 
of the histogram represents the 17 objects for which we have identified 
IRAC counterparts to the submillimeter sources.
The black portion of the histogram includes only those five sources
we call AGN-dominated on the basis of the IRAC-MIPS SEDs alone (14.3, 14.7, 14.12, 14.13, 
and 14.19).  
\label{fig:fluxhist} }
\end{figure}

\begin{figure}[!t]
\plotone{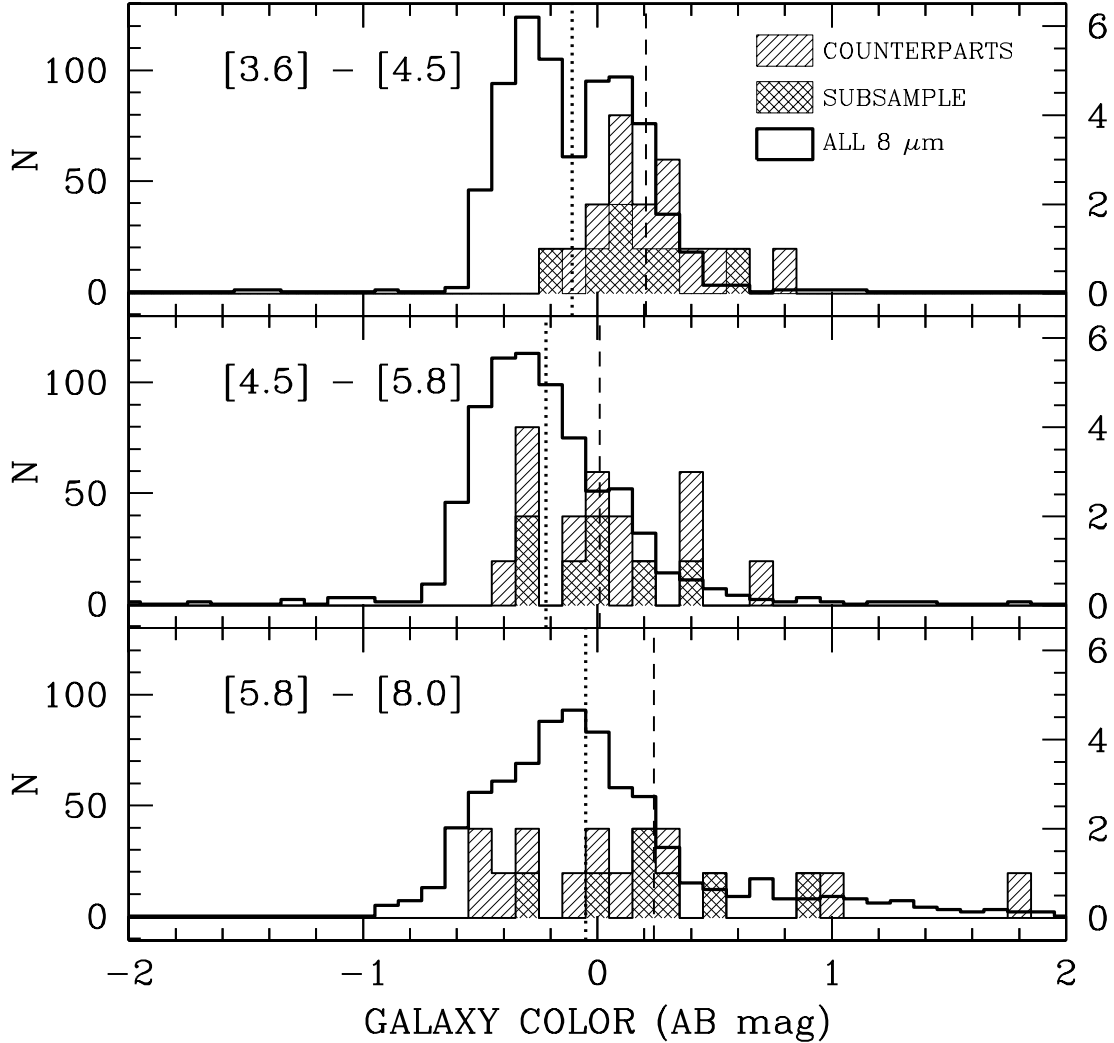}
\caption{Histograms showing the distributions of IRAC colors for the
submillimeter galaxies' counterparts relative to those of
all 8$\mu$m-detected galaxies in the 14$^{\rm h}$ field.  
Top panel: $[3.6]-[4.5]$ color.  The 17 IRAC counterparts lie
under the shaded histogram and are referenced to the right-hand
vertical axis.  The field galaxy sample lies under the unshaded histogram
and is referenced to the left-hand vertical axis.  The dotted
and dashed lines indicate the mean colors for the all 8~$\mu$m-detected
galaxies and the counterpart sample, respectively.  Center and bottom
panels show the distributions of the
$[4.5]-[5.8]$ and $[5.8]-[8.0]$ colors, respectively.
In all panels the doubly-hatched histogram indicates the
distribution of a counterpart subsample consisting of
the seven objects for which no color criteria were used
to identify a counterpart.
\label{fig:histocolor} }
\end{figure}

\begin{figure}[!t]
\plotone{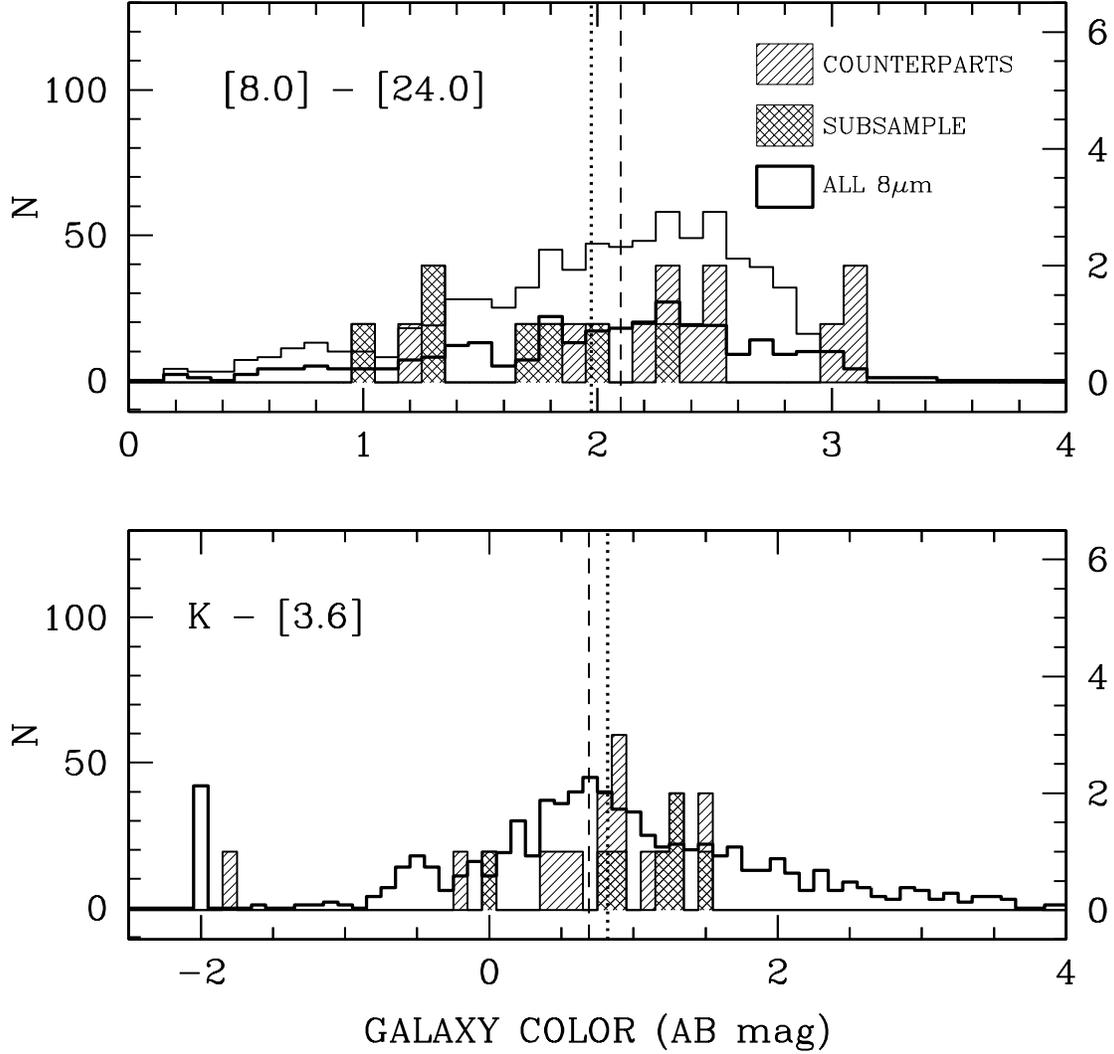}
\caption{Upper panel: The distributions of $[8.0] - [24]$ 
colors for the IRAC-selected counterparts in the 14$^{\rm h}$ field 
(shaded histograms, referenced to the right-hand vertical axis) 
and those of the full sample of all 8$\mu$m-detected IRAC galaxies.
The latter lie
under two unshaded histograms referenced to the left-hand axis; the lower 
histogram indicates the distribution only for galaxies with significant MIPS
24$\mu$m detections.  The upper unshaded histogram represents all
8$\mu$m-detected galaxies but uses 24~$\mu$m upper limits where
detections are unavailable.
The counterpart subsample under the doubly-hatched histogram consists
of the seven objects for which no color criteria were used
to identify a counterpart.
The dotted and dashed lines indicate the mean colors for the full sample
and the counterpart sample, respectively. 
Lower panel: Symbols as for the upper panel, but showing the 
distributions of {\sl K} - [24] colors for the IRAC-selected 
counterparts (shaded histograms) and those of
the full sample of 8~$\mu$m-detected galaxies (unshaded histogram).
\label{fig:mipscolor} }
\end{figure}

\clearpage

\begin{deluxetable}{lcrrrrrccll}
\tabletypesize{\scriptsize}
\tablecaption{All Candidate IRAC 8.0~$\mu$m Counterparts to SCUBA Sources \label{8micids}}
\tablewidth{0pt}
\tablehead{
  \colhead{SCUBA} & 
  \colhead{Candidate} & 
  \colhead{F$_{3.6}$} & 
  \colhead{F$_{4.5}$} & 
  \colhead{F$_{5.8}$} & 
  \colhead{F$_{8.0}$} & 
  \colhead{F$_{24.0}$} & 
  \colhead{R.A.} & 
  \colhead{Decl.} & 
  \colhead{Offset\tablenotemark{a}} & 
  \colhead{P$^\prime$} \\ 
  \colhead{Name} & 
  \colhead{} & 
  \colhead{($\mu$Jy)} & 
  \colhead{($\mu$Jy)} & 
  \colhead{($\mu$Jy)} & 
  \colhead{($\mu$Jy)} & 
  \colhead{($\mu$Jy)} & 
  \multicolumn{2}{c}{(J2000)} &
  \colhead{(arcsec)} & 
  \colhead{} 
}
\startdata
 $^\dagger$14.1\tablenotemark{b}
        & &     40.7 &   44.8 &   32.5 &   41.1 &  191   & 14 17 40.22 & 52 29 06.5  & 1.07(0.28) & 0.11 \\
 $^\dagger$14.2   
        &A&     38.5 &   45.0 &   41.5 &   27.0 &  150   & 14 17 51.33 & 52 30 24.9  & 6.55 & 0.18 \\
 14.2   &B&      2.0 &    3.3 &    5.5 &    8.6 &  $<70$ & 14 17 51.93 & 52 30 30.8  & 2.12 & 0.45 \\
 $^\dagger$14.3\tablenotemark{b}
        & &     35.0 &   41.0 &   48.6 &   48.4 &  408   & 14 18 00.58 & 52 28 21.3  & 0.87(2.32) & 0.078 \\
 $^\dagger$14.4   
        &A&      2.7 &    3.6 &    5.2 &   27.2 &  239   & 14 17 43.43 & 52 28 05.8  & 8.71 & 0.18 \\
 14.4   &B&     52.8 &   49.2 &   40.7 &   40.9 &\nodata & 14 17 42.41 & 52 28 11.6  & 9.07 & 0.11 \\
 14.4   &C&      7.4 &    8.7 &    9.5 &    6.2 &  $<70$ & 14 17 43.80 & 52 28 16.8  & 4.71 & 0.52 \\
 14.6   & &      1.8 &    1.8 &  $<6.3$&    5.5 &  $<70$ & 14 17 57.02 & 52 29 14.7  & 8.50 & 0.51 \\
 14.7   &A&      8.4 &   10.6 &   11.7 &   12.9 &  $<70$ & 14 18 00.95 & 52 29 50.0  & 1.68 & 0.35 \\
 $^\dagger$14.7   
        &B&      5.3 &   10.6 &   20.2 &   51.3 &  150   & 14 18 01.15 & 52 29 42.0  & 7.02 & 0.071 \\
 $^\dagger$14.8   
        &A&     21.4 &   24.5 &   24.7 &   15.3 &  275   & 14 18 02.01 & 52 30 16.3  & 6.43 & 0.31 \\
 14.8   &B&      5.4 &    5.6 &    4.9 &    2.9 &  $<70$ & 14 18 02.70 & 52 30 10.4  & 5.25 & 0.76 \\
$^\dagger$14.10   
        &A&     50.1 &   48.4 &   34.8 &   38.3 &  307   & 14 18 03.00 & 52 29 33.2  & 9.81 & 0.12 \\
14.10   &B&      4.3 &    3.2 &  $<6.3$&  $<5.8$&  $<70$ & 14 18 03.98 & 52 29 40.6  & 2.20 & 0.86 \\
14.11   &A&    153.6 &   99.5 &   66.7 &   36.0 &  $<70$ & 14 17 46.89 & 52 32 36.1  & 2.69 & 0.13 \\
14.11   &B&    225.3 &  140.8 &   92.7 &   50.3 &  $<70$ & 14 17 47.32 & 52 32 42.1  & 4.60 & 0.075 \\
$^\dagger$14.12   
        &A&      3.7 &    5.7 &  $<6.3$&    8.0 &   80   & 14 18 04.70 & 52 28 56.0  & 5.48 & 0.47 \\
14.12   &B&      3.1 &    4.2 &  $<6.3$&   18.3 &  $<70$ & 14 18 05.69 & 52 29 03.3  & 8.60 & 0.33 \\
$^\dagger$14.13\tablenotemark{b}
        & &    580.1 &  981.7 & 1448.5 & 2225.5 & 5646   & 14 17 41.88 & 52 28 23.5  & 0.67(6.46) & 0.0013 \\
14.14   &A&      9.9 &    8.3 &    3.4 &    6.2 &  $<70$ & 14 18 09.06 & 52 31 00.5  & 4.78 & 0.52 \\
$^\dagger$14.14   
        &B&      9.4 &   10.7 &    7.8 &    6.9 &  120   & 14 18 08.21 & 52 31 08.1  & 6.16 & 0.50 \\
14.15   &A&     30.8 &   26.7 &   19.7 &   17.1 &  103   & 14 17 29.87 & 52 28 21.5  & 5.83 & 0.28 \\
$^\dagger$14.15   
        &B&      7.4 &   10.8 &   15.3 &   12.1 &   93   & 14 17 28.44 & 52 28 19.8  & 7.90 & 0.37 \\
$^\dagger$14.17   & &     47.1 &   47.0 &   34.6 &   77.0 &  266   & 14 17 24.44 & 52 30 46.1  & 9.48 & 0.049 \\
$^\dagger$14.18\tablenotemark{b}
        & &    159.7 &  138.1 &  143.0 &  164.7 & 1078   & 14 17 42.11 & 52 30 25.7  & 0.61(1.51) & 0.029 \\
$^\dagger$14.19\tablenotemark{c}
        & &     15.3 &   20.6 &   21.1 &   25.2 &   84.0 & 14 18 11.26 & 52 30 12.3  & 2.33(8.59) & 0.20 \\
$^\dagger$14.20   
        & &     18.5 &   21.2 &   18.6 &   13.9 &  $<70$ & 14 17 50.51 & 52 30 55.2  & 8.87 & 0.34 \\
14.22   &A&     29.2 &   28.1 &   19.6 &   19.1 &  $<70$ & 14 17 55.08 & 52 32 08.4  & 7.91 & 0.25 \\
$^\dagger$14.22   
        &B&     72.1 &   66.7 &   49.1 &   50.3 &  500   & 14 17 56.80 & 52 31 57.8  & 9.94 & 0.075 \\
$^\dagger$14.23   
        &A&     31.7 &   43.4 &   47.1 &   34.0 &  529   & 14 17 46.21 & 52 33 22.2  & 1.97 & 0.14 \\
14.23   &B&     21.4 &   17.5 &   20.1 &   20.1 &  $<70$ & 14 17 46.68 & 52 33 29.6  & 6.60 & 0.24
\enddata
\tablenotetext{a}{Offsets to SCUBA positions given in parentheses where 1.4 GHz data are available.}
\tablenotetext{b}{Counterpart search based on 1.4 GHz detections from Eales \etal (2000).}
\tablenotetext{c}{Counterpart search based on 1.4 GHz position reported by Webb \etal (2003).}
\tablecomments{ The best 8~$\mu$m candidates are indicated with $\dagger$ (see text).
Flux density upper limits given are 5$\sigma$ point source 
sensitivities.  Positions are measured in the IRAC 8.0~$\mu$m mosaic and have 
roughly 0\farcs3 uncertainties.
Units of right ascension are hours, minutes, and seconds, and units of declination
are degrees, arcminutes, and arcseconds.  
}
\end{deluxetable}

\begin{deluxetable}{lcccccccccc}
\tabletypesize{\scriptsize}
\tablecaption{Ground-Based Data for Best 8.0~$\mu$m Counterpart Candidates \label{groundphot}}
\tablewidth{0pt}
\tablehead{
  \colhead{Candidate} & 
  \colhead{U\tablenotemark{a}} & 
  \colhead{B\tablenotemark{a}} & 
  \colhead{V\tablenotemark{a}} & 
  \colhead{R\tablenotemark{b}} & 
  \colhead{I\tablenotemark{a}} & 
  \colhead{z\tablenotemark{c}} & 
  \colhead{K\tablenotemark{d}} &  
  \colhead{Redshift} 
}
\startdata
 14.1 & 27.17$\pm$0.32 & 26.60$\pm$0.12 & 26.28$\pm$0.11 & 26.6$\pm$0.1  & 24.71$\pm$0.05 & 24.79$\pm$0.34 & 21.18$\pm$0.05 & \nodata \\
 14.2A& 28.0$\pm$0.6   & 25.9$\pm$0.1   & 26.64$\pm$0.07 & 24.94$\pm$0.06 & 25.0$\pm$0.05 & \nodata        & 21.4$\pm$0.1   & \nodata \\
 14.3 & 24.71$\pm$0.06 & 24.55$\pm$0.05 & 24.06$\pm$0.05 & 24.97$\pm$0.06 & 23.20$\pm$0.05& 22.78$\pm$0.05 & 21.23$\pm$0.05 & \nodata \\
 14.4A& 22.71$\pm$0.05 & 22.60$\pm$0.05 & 22.70$\pm$0.05 & 21.74$\pm$0.02& 22.37$\pm$0.05 & $>25.0$        & $>21.0$        & \nodata \\
 14.7B& 24.53$\pm$0.06 & 23.64$\pm$0.05 & 23.41$\pm$0.05 & 23.5$\pm$0.05 & 23.49$\pm$0.05 & \nodata        & 22.5$\pm$0.2   & \nodata \\
 14.8A& 25.93$\pm$0.13 & 25.48$\pm$0.05 & 24.96$\pm$0.05 & 25.0$\pm$0.1  & 24.58$\pm$0.05 & 24.24$\pm$0.20 & 21.5$\pm$0.1   & \tablenotemark{e} \\
14.10A& 27.29$\pm$0.34 & 26.74$\pm$0.13 & 27.01$\pm$0.23 & 25.7$\pm$0.1  & 24.65$\pm$0.05 & 23.83$\pm$0.14 & 20.8$\pm$0.1   & \nodata \\
14.12A& $>27.71$       & $>26.23$       & 27.60$\pm$0.37 & 26.4$\pm$0.2  & 25.54$\pm$0.09 & $>25.0$        & $>23.1$        & \nodata \\
14.13 & 23.93$\pm$0.05 & 23.73$\pm$0.05 & 22.90$\pm$0.05 & 22.02$\pm$0.02 & 20.86$\pm$0.05 & 20.39$\pm$0.05 & 18.5$\pm$0.1   & 1.150 \\
14.14B& 24.80$\pm$0.07 & 24.32$\pm$0.02 & 24.02$\pm$0.02 & 23.76$\pm$0.07& 23.53$\pm$0.02 & 22.99$\pm$0.11 & 22.0$\pm$0.2   & \nodata \\
14.15B&  $>27.71$      & $>26.23$       & $>25.98$       & $>26.6$         & 25.13$\pm$0.06 & \nodata        & $>21.5$        & \nodata \\
14.17 & 23.84$\pm$0.05 & 23.11$\pm$0.05 & 22.29$\pm$0.05 & 21.69$\pm$0.02& 21.13$\pm$0.05 & 21.00$\pm$0.05 & 19.7$\pm$0.1   & \nodata \\
14.18 & 22.97$\pm$0.05 & 22.57$\pm$0.05 & 21.99$\pm$0.05 & 21.36$\pm$0.02& 20.61$\pm$0.05 & 20.56$\pm$0.05 & 19.7$\pm$0.1   & 0.661 \\
14.19 & 26.57$\pm$0.19 & 24.95$\pm$0.03 & 24.36$\pm$0.03 & 24.1$\pm$0.04 & 24.08$\pm$0.05 & 24.10$\pm$0.18 & 21.7$\pm$0.1   & \nodata \\
14.20 & $>27.71$       & $>26.23$       & $>25.98$       & 27.7$\pm0.6$  & 25.63$\pm$0.65 & 25.83$\pm$0.88 & 21.6$\pm$0.1   & \tablenotemark{e} \\
14.22B& $>27.71$       & $>26.23$       & $>25.98$       & 24.83$\pm$0.07& 23.55$\pm$0.1  &        \nodata & 20.19$\pm$0.1  & \nodata \\

14.23A& $>27.71$       & 26.75$\pm$0.15 & 26.57$\pm$0.16 & 27.7$\pm$0.6  & 25.09$\pm$0.07 & 25.53$\pm$0.67 & 20.9$\pm$0.2   & \nodata 
\enddata
\tablecomments{All magnitudes are given on the AB system.  Sources of the data: $^a$ CFDF (McCracken
\etal 2001); $^b$ Miyazaki (2006, private communication); $^c$ Brodwin \etal (2006); $^d$ K-band survey of Webb \etal 
(2003); $^e$ Chapman \etal (2005) report a spectroscopic $z=2.128$ for galaxies near but distinct from 
these positions.  
}  
\end{deluxetable}
\label{tab:photometry}

\begin{deluxetable}{lrrrrrrr}
\tabletypesize{\scriptsize}
\tablecaption{Mean Infrared Colors of IRAC-Selected Galaxies in the CUDSS 14$^{\rm h}$ Field. \label{tab:colors}}
\tablewidth{0pt}
\tablehead{
  \colhead{Sample} &
  \colhead{N} &
  \colhead{K-[3.6]} &
  \colhead{[3.6]-[4.5]} &
  \colhead{[4.5]-[5.8]} &
  \colhead{[5.8]-[8.0]} &
  \colhead{[8.0]-[24]} \\
}
\startdata
8~$\mu$m Galaxies      & 726 &  $0.8\pm1.2$ & $-0.1\pm0.4$ &  $-0.1\pm0.6$  &   $0.0\pm0.5$ & $1.9\pm0.7$\\ 
8~$\mu$m Counterparts  &  17 &  $0.7\pm0.8$ &  $0.2\pm0.2$ &   $0.0\pm0.3$  &   $0.2\pm0.6$ & $2.1\pm0.7$\\
Counterparts Subsample &   7 &  $1.0\pm0.5$ &  $0.2\pm0.2$ &   $0.0\pm0.3$  &   $0.2\pm0.4$ & $1.6\pm0.4$ 
\enddata
\end{deluxetable}


\clearpage

\begin{thebibliography}{}


\bibitem[Alexander \etal (2004)]{ale04}
Alexander, D.~M., Smail, I., Bauer, F.~E., Chapman, S.~C., Blain, A.~W., 
Brandt, W.~N., \& Ivison, R.~J.\ 2004, \nat, 434, 738

\bibitem[Barger \etal (1988)]{bar98}
Barger, A.~J., et al. 1998, Nature, 394, 428

\bibitem[Barmby {et~al.}(2006)]{bar06}
Barmby, P. {et~al.} 2006, \apj, in press (astro-ph/0512618).

\bibitem[Bertin \& Arnouts (1996)]{sextractor}
Bertin, E., \& Arnouts, S., 1996, \aaps, 117, 393 

\bibitem[Blain et al.(2002)]{2002PhR...369..111B} Blain, A.~W., Smail, I., 
Ivison, R.~J., Kneib, J.-P., \& Frayer, D.~T.\ 2002, \physrep, 369, 111 

\bibitem[Borys et al.(2004)]{bor04} 
Borys, C., et al.\ 2005, MNRAS, 355, 485

\bibitem[Brodwin et al.(2006)]{bro05} 
Brodwin, M., et al.\ 2006, ApJS, 162, 20

\bibitem[Chapman et al.(2003)]{2003Natur.422..695C} Chapman, S.~C., Blain, 
A.~W., Ivison, R.~J., \& Smail, I.~R.\ 2003, \nat, 422, 695 

\bibitem[Chapman et al.(2005)]{chapman05} Chapman, S.~C., Blain, 
A.~W., Smail, I.~R., \& Ivison, R.~J.\ 2005, \apj, 622, 722


\bibitem[Clements et al.(2004)]{2004MNRAS.351..447C} Clements, D., et al.\ 
2004, \mnras, 351, 447 

\bibitem[Cole et al.(2001)]{2001MNRAS.326..255C} Cole, S., et al.\ 2001, 
\mnras, 326, 255 

\bibitem[Dye et al.(2006)]{dye06} Dye, S., et al.\ 2006, \apj, accepted 
(astro-ph/0512357).

\bibitem[Eales et al.(1999)]{1999ApJ...515..518E} Eales, S., Lilly, S., 
Gear, W., Dunne, L., Bond, J.~R., Hammer, F., Le F{\` e}vre, O., \& 
Crampton, D.\ 1999, \apj, 515, 518 

\bibitem[Eales et al.(2000)]{eales} Eales, S., Lilly, S., Webb, T., Dunne, L.,    
Gear, W., Clements, D., \& Yun, M. 2000, \aj, 120 2244

\bibitem[Egami et al.(2004)]{2004ApJS..154..130E} Egami, E., et al.\ 2004, 
\apjs, 154, 130 

\bibitem[Fazio et al.(2004)]{irac} Fazio, G. G., et al. 2004, \apjs, 154, 10   

\bibitem[Flores et al.(1999)]{1999ApJ...517..148} Flores, H., \etal 1999, \apj, 517, 148

\bibitem[Fomalont et al.(1991)]{fom91}
Fomalont, E.~B., Windhorst, R.~A., Kristian, J.~A., \& Kellerman, K.~I.\ 1991, \aj, 102, 1258


\bibitem[Frayer et al.(2004a)]{2004AJ....127..728F} Frayer, D.~T., et al.\ 
2004a, \aj, 127, 728 

\bibitem[Frayer et al.(2004b)]{2004ApJS..154..137F} Frayer, D.~T., et al.\ 
2004b, \apjs, 154, 137 

\bibitem[Gordon et al. (2004)]{mipsdata}
Gordon, K.~S., et al., 2004, SPIE vol. 5487, ed. J. C. Mather, 177

\bibitem[{{Hatziminaoglou} {et~al.}(2005)}]{hatz05}
{Hatziminaoglou}, E. {et~al.} 2005, \aj, 129, 1198

\bibitem[Higdon et al.(2004)]{higdon} Higdon, S. J. U., et al. 2004, \apjs, 154, 174   

\bibitem[Huang et al.(2004)]{2004ApJS..154...44H} Huang, J.-S., et al.\ 
2004, \apjs, 154, 44 

\bibitem[Hughes et al.(1998)]{hug98} Hughes, D.~H., et al. 1998,
Nature, 394, 241

\bibitem[Ivison et al.(2002)]{ivi02} Ivison, R.~J., et al.\ 
2002, \mnras, 337, 1  


\bibitem[Knudsen et al.~(2006)]{knu06} 
Knudsen, K.~K., et al.\ 2006, MNRAS, in press (astro-ph/0602131)

\bibitem[{{Lacy} {et~al.}(2004)}]{lacy04}
{Lacy}, M. {et~al.} 2004, \apjs, 154, 166

\bibitem[Lilly et al.(1999)]{lilly} Lilly, S.J., Eales, S. A., Gear, K.P., 
Hammer, F., Le F\`evre, O., Crampton, D., Bond, J.R., and Dunne, L. 1999,
\apj, 518, 641

\bibitem[McCracken et al. (2001)]{cfdf}
McCracken, H.~J., Le F{\` e}vre, O., Brodwin, M., Foucaud, S., 
Lilly, S.~J., Crampton, D., \& Mellier, Y., 2001 \aa, 376, 756


\bibitem[Nandra et al.\ (2005)]{2005MNRAS.356..568N} Nandra, K., et al.\ 
2005, \mnras, 356, 568 

\bibitem[Pope et al.(2005)]{2005MNRAS.358..149P} Pope, A., Borys, C., 
Scott, D., Conselice, C., Dickinson, M., \& Mobasher, B.\ 2005, \mnras, 
358, 149 

\bibitem[Rieke et al.\ (2004)]{mips} Rieke, G., et al. 2004, \apjs, 154, 25   


\bibitem[Schmitt et~al. (2006)]{schmitt06}
{Schmitt}, H.~R. {et~al.} 2006, in preparation (astro-ph/0602064)

\bibitem[Serjeant et al. (2004)]{ser04}
Serjeant, S., et al.\ 2004, \apjs, 154, 118

\bibitem[Smail, Ivison, \& Blain (1997)]{sma97} 
Smail, I., Ivison, R.~J., \& Blain, A.~W. 1997, \apj, 490, L5

\bibitem[Smail, \etal (2004)]{sma04} 
Smail, I., Chapman, S.~C., Blain, A.~W., \& Ivison, R.~J. 2004, \apj, 616, 71

\bibitem[{{Stern} {et~al.}(2005)}]{stern05}
{Stern}, D. {et~al.} 2005, \apj, 631, 163


\bibitem[Wang, Cowie, \& Barger (2004)]{wang04} 
Wang, W.-H., Cowie, L.~L., and Barger, A.~J. 2004, \apj, 655, 671

\bibitem[Waskett et al.(2003)]{2003MNRAS.341.1217W} 
Waskett, T.~J., et al.\ 2003, \mnras, 341, 1217 

\bibitem[Webb et al.(2003)]{webb} 
Webb, T. M. A., et al.\ 2003, \apj, 597, 680


\bibitem[Yun \& Carilli (2002)]{yuncar:02}
Yun, M.~S., \& Carilli, C.~L., 2002, \apj, 568, 2002

\end{thebibliography}
\end{document}